\documentclass[pdflatex,sn-basic]{sn-jnl}% Basic Springer Nature Reference Style/Chemistry Reference Style
\usepackage{graphicx}%
\usepackage{multirow}%
\usepackage{amsmath,amssymb,amsfonts}%
\usepackage{amsthm}%
\usepackage{mathrsfs}%
\usepackage[title]{appendix}%
\usepackage{xcolor}%
\usepackage{textcomp}%
\usepackage{manyfoot}%
\usepackage{booktabs}%
\usepackage{algorithm}%
\usepackage{algorithmicx}%
\usepackage{algpseudocode}%
\usepackage{listings}%
\usepackage{textcomp}

\usepackage{bm}
\usepackage{ulem}

\usepackage{array} 

\theoremstyle{thmstyleone}%
\theoremstyle{thmstyletwo}%

\theoremstyle{thmstylethree}%

\begin{document}

\title[Unified Deflection Estimation and Error Analysis for Background-Oriented Schlieren]{Unified Deflection Estimation and Error Analysis for Background-Oriented Schlieren}

\author[1]{\fnm{Jiawei} \sur{Li}} %\email{lijiawei@buaa.edu.cn}
\author[1]{\fnm{Xiang} \sur{Li}} %\email{x.li@buaa.edu.cn}
\author[1]{\fnm{Chong} \sur{Pan}} %\email{iiauthor@gmail.com}
\author*[1,2]{\fnm{Yuan} \sur{Xiong}}\email{xiongyuan@buaa.edu.cn}

\affil*[1]{\orgdiv{Institute of Fluid Mechanics \& Aircraft and Propulsion Laboratory, Ningbo Institute of Technology}, \orgname{Beihang University}, \orgaddress{\street{37, Xueyuan Road}, \city{HaiDian District}, \postcode{100191}, \state{Beijing}, \country{China}}}

\affil*[2]{\orgname{Tianmushan Laboratory}, \orgaddress{\street{166 Shuanghongqiao Street}, \city{Hangzhou}, \postcode{315800}, \state{Zhejiang}, \country{China}}}


\abstract{
Background-Oriented Schlieren (BOS) has become a versatile quantitative diagnostic for density-varying flows, in which estimating the light-ray deflection from the measured displacement is the essential step linking the recorded images to the underlying refractive-index field. Two-dimensional BOS traditionally treats this through the intuitive deflection angle, whereas three-dimensional tomographic BOS relies on the rigorous deflection vector derived from the ray equation. These descriptions have evolved largely independently, and the assumptions bridging them, together with the systematic errors they introduce, have not been examined in a unified manner. Based on geometric optics, this study establishes a unified deflection estimation framework that reconciles the mainstream two- and three-dimensional methods into a single mathematical structure and exposes the hierarchy of approximations underlying each. By deconstructing four key assumptions, namely the thin phase object, the uniform boundary refractive index, the paraxial approximation, and the perpendicularity between the deflection vector and the optical axis, we derive rigorous unified deflection expressions in both two- and three-dimensional space and categorize the mainstream methods accordingly. Using phase objects constructed from one-dimensional chirp signals and two-dimensional turbulent fields from Direct Numerical Simulation, combined with high-fidelity nonlinear ray tracing as the ground truth, we quantitatively characterize and analytically interpret the deflection estimation error of each method under both uniform and non-uniform refractive-index boundary conditions. This work provides a theoretical toolkit for assessing and enhancing the accuracy of quantitative BOS diagnostics.
}




\maketitle

\section{Introduction}\label{sec_Intro}

Non-uniform-density fluids are widely encountered in natural phenomena and engineering applications. Visualizing the associated density field is critical for understanding the underlying flow physics. For example, in high-speed aerodynamics, the flow density field directly indicates the structure of shock and compression waves \citep{venkatakrishnan_density_2004,lin_study_2019, heineck_background-oriented_2021, leon_three-dimensional_2022, li_experimental_2024, yang_shock_2024}. Instead of directly probing the flow density field $\rho$, the refractive index $n$ is typically measured owing to the Gladstone–Dale (G-D) relationship $n-1=G\rho$ where $G$ is the G-D constant depending on the gas composition and light wavelength. For decades, knife-edge-based Schlieren techniques have been the prevailing optical methods for probing density-varying flows, where light deflections are converted into changes in image intensity \citep{settles_review_2017}. However, classical methods suffer from tedious optical alignment, in-situ quantitative calibration, and a limited field of view (FOV) \citep{hargather_comparison_2012,settles_review_2017}. 

Background-Oriented Schlieren (BOS), as a new variant of the Schlieren technique, was proposed around 2000 \citep{richard_principle_2001, dalziel_whole-field_2000, meier_computerized_2002}. Compared to the classical Schlieren, BOS is advantageous in terms of a simple optical setup, offline geometric calibration, and a large FOV \citep{raffel_background-oriented_2015, schmidt_twenty-five_2025}. Owing to the well-developed cross-correlation algorithm in the particle image velocimetry community, BOS can achieve sub-pixel accuracy in estimating the image displacements combined with a randomly dotted background pattern of the image pair with/without the phase object (PO) \citep{dalziel_whole-field_2000, roesgen_optimal_2003, westerweel_particle_2013}. To overcome the algorithmic window size limitation, optical flow methods \citep{atcheson_evaluation_2009, schmidt_wavelet-based_2021}, the single-pixel correlation method \citep{sugisaki_single-pixel_2022}, and Fourier demodulation-based algorithms \citep{wildeman_real-time_2018, shimazaki_background_2022} are developed, and the spatial resolution of the displacement estimations is improved significantly while retaining the sub-pixel accuracy \citep{schmidt_twenty-five_2025}.

Given a satisfactory displacement field, the light deflections induced by PO can subsequently be estimated. However, such an estimation is not as trivial as expected. For quasi-two-dimensional or axisymmetric flows, the light deflections are widely interpreted as the ‘deflection angle’ subtended by the incident and refracted rays \citep{richard_principle_2001, venkatakrishnan_density_2004, van_hinsberg_density_2014, xiong_towards_2020}. Utilizing the geometric trigonometric relation and combined with the paraxial assumption, such deflection angle estimation can be routinely expressed in a neat form as the displacement distance divided by the distance between the background and the center plane of PO \citep{raffel_background-oriented_2015}, the length of which can be obtained by easy geometric length calibrations. Following the trigonometric approach, extensive efforts are made to relax the paraxial assumption by accounting for the angle between the light ray and the optical axis; thus, the deflection angle can be approximated more accurately \citep{goldhahn_background_2007}. However, the deflection angle concept implicitly assumes that the background plane is perpendicular to the optical axis, and its detailed relationship to the ray equation is rarely elucidated.

To measure an $n$ field with arbitrary spatial distributions, the tomographic background oriented Schlieren (TBOS) concept with a multi-view setup was introduced only a few years after the appearance of BOS and has been applied for various applications \citep{atcheson_time-resolved_2008, nicolas_direct_2016, nicolas_3d_2017, lang_measurement_2017, grauer_instantaneous_2018, grauer_fast_2020, liu_volumetric_2021, bo_background-oriented_2023, hu_reconstruction_2024, jia_tomographic_2024, li_three-dimensional_2024, li_neural_2024}. Using tomographic algorithms such as algebraic reconstruction techniques, simultaneous iterative reconstruction techniques, and conjugate gradient least-squares, TBOS reconstructs the 3D $n$ field from displacement fields across multiple views. Estimating the light deflections with TBOS can be categorized into two groups. One is a direct extension of the deflection angle concept, which omits the light deflection in the principal optical axial direction \citep{grauer_instantaneous_2018, amjad_assessment_2020, liu_volumetric_2021, bo_background-oriented_2023}. Consequently, the estimation error can become significant when the background plane is skewed with respect to the optical axis. The other group relies on a rigorous derivation of the light-deflection vector, with three components, from the ray equation \citep{atcheson_time-resolved_2008, nicolas_direct_2016, lang_measurement_2017, li_three-dimensional_2024}. To simplify the deflection estimation, various assumptions are further involved, such as the constant $n_0$ at all boundaries of the reconstructed volume, a fixed refracted plane for all rays, and a pinhole lens model \citep{atcheson_time-resolved_2008, nicolas_direct_2016}. These assumptions inevitably introduce errors in deflection estimation. However, to the authors' knowledge, a systematic clarification of deflection estimation in two- or three-dimensional BOS measurements is still lacking, as is an analysis of associated errors. 

In this regard, this paper aims to theoretically establish a unified framework for deflection estimation that encompasses popular methods and to analyze the associated errors and assumptions. Section~\ref{sec_Theory} of this paper summarizes the popular deflection estimation methods in the BOS community and proposes a unified deflection estimation framework. Section~\ref{sec:RayTracing} presents the nonlinear ray-tracing simulation platform and the specifically designed flow fields involved. Section~\ref{sec_Results} presents the associated error analysis for the deflection estimations. Finally, Section~\ref{sec_Concl} summarizes the conclusions.

%%%%%%%%%%%%%%%%%%%%%%%%%%%%%%%%%%%%%%%%%%%%%%%%%%%%%
%%%%%%%%%%%%%%%%%%%%%%%%%%%%%%%%%%%%%%%%%%%%%%%%%%%%%
%%%%%%%%%%%%%%%%%%%%%%%%%%%%%%%%%%%%%%%%%%%%%%%%%%%%%

\section{Deflection Estimation Theory}\label{sec_Theory}
\begin{figure}[htb]
\centering
\includegraphics[width=0.7\textwidth]{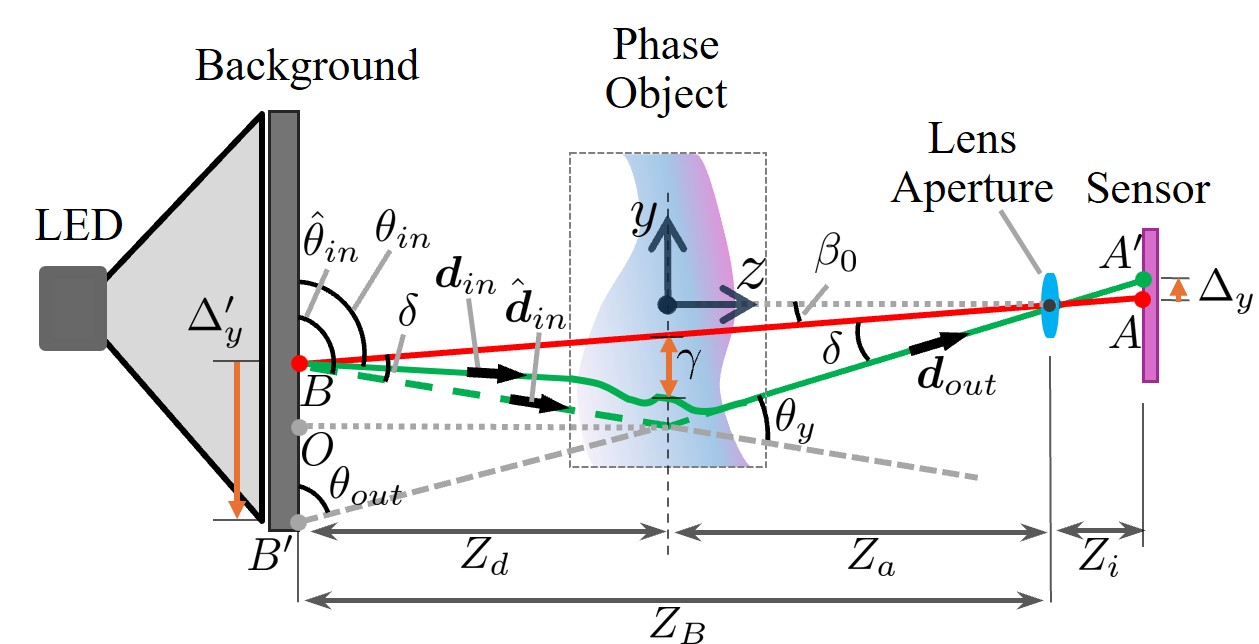}
\caption{Schematic of a canonical BOS experimental setup.}\label{fig-BOS-Schematic}
\end{figure}

A canonical setup for 2D BOS typically comprises a camera, a background with random dots, and a light source to illuminate the background, as schematically shown in Fig.~\ref{fig-BOS-Schematic}. The target flow of a non-uniform $n$ is allocated between the camera lens and the background. Usually, the BOS camera focuses on the background pattern, and background images both with and without a PO are recorded. As shown in Fig.~\ref{fig-BOS-Schematic}, assuming a pinhole camera model, the point $B$ originally forms the image $A$ on the sensor plane without the PO, following the solid red ray trajectory. With PO present, in order to enter the pinhole, another ray trajectory emitted by $B$ has to be selected as indicated by the green solid line. Typically, a PO is of finite thickness; thus, the light ray propagates with a nonlinear trajectory, as exaggerated by the path wrinkles on the solid green line. This disturbed light ray from $B$ arrives at $A^\prime$ on the sensor plane with corresponding virtual image $B^\prime$. The displacement between $B$ and $B^\prime$ is $\Delta_y^\prime$. 

\subsection{2D BOS Deflection Estimation}\label{subsec:BOS_principle}

Intuitively, the light deflection is interpreted as the angle subtended by the incident and refracted ray directional vector ${\bm{d}}_{in}$ and $\bm{d}_{out}$ (assuming 2D scenario) in the schematic, where subscripts "in" and "out" indicate the ray entering and exiting the flow field, respectively. While $\bm{d}_{out}$ can be determined with known pixel displacement on the sensor plane with the pinhole model, $\bm{d}_{in}$ remains unknown due to the nonlinear ray trajectory inside PO. A widely accepted assumption is that a fixed deflection plane exists at which all ray deflections occur. For convenience, the deflection plane is usually allocated at the center or boundaries of the PO, equivalent to a thin PO assumption  \citep{atcheson_time-resolved_2008, nicolas_direct_2016}. Consequently, $\bm{d}_{in}$ can be replaced with  $\hat{\bm{d}}_{in}$ thus angle $\theta_y$ can be easily determined. According to Fig.~\ref{fig-BOS-Schematic}, a detailed geometric trigonometric form of $\theta_y$ can be expressed as following \citep{goldhahn_background_2007},           
\begin{equation}
\tan\theta_y = \frac{(1+\frac{Z_{d}}{Z_{a}})\Delta_y\frac{Z_{B}-f}{Z_{B} f} \cos^2\beta_0}{\frac{Z_{d}}{Z_{a}}-(\Delta_y\frac{Z_{B}-f}{Z_{B}  f}\cos^2\beta_0)^2}.
\label{eq_deflection_geometric_detailed}
\end{equation}
It is claimed that if $Z_B \gg f$ and paraxial assumptions are applied, such a deflection angle estimation can be simplified in a widely adopted neat form as, 
\begin{equation}
\theta_y \approx \Delta_y^\prime/Z_d,
\label{eq:deflection_geometric}
\end{equation}
where $\Delta_y^\prime$ is the displacement distance and $Z_d$ is the distance between the background and the center plane of the PO. These lengths can be easily obtained through length calibrations. On the other hand, $\theta_y$ can also be linked to $n$ approximately as:
\begin{equation}
\theta_y\approx\frac{1}{n_0}\int \frac{\partial n}{\partial y} dz.
\label{eq_epsilonDefine}
\end{equation}
Considering both $\theta_x$ and $\theta_y$ for a nominal two-dimensional PO, tackling the BOS problem is equivalent to solving a Poisson equation with proper boundary conditions (BCs), as shown by \cite{xiong_towards_2020, xiong_analysis_2020}. Once an $n$ field is obtained, flow density $\rho$ can be inferred owing to the G-D relation. More details regarding the canonical 2D BOS methodology can be found in the aforementioned BOS reviews \citep{richard_principle_2001, raffel_background-oriented_2015, schmidt_twenty-five_2025}.

\subsection{Tomographic BOS Deflection Estimation}\label{subsec:TBOS_principle}
While describing light deflection as the angle subtended by the incident and refracted rays is physically sound for two-dimensional BOS, extending this description to tomographic BOS (TBOS) is nontrivial. Begin with the ray equation governing the trajectory of a light ray within the $n$ field:
\begin{equation}
\frac{d}{ds}(n \frac{d\bm{r}}{ds})=\nabla{n},
\label{eq_RayEq}
\end{equation}
where $\bm{r}$ is the position vector along the ray path, $ds$ denotes the infinitesimal path element, and $d\bm{r}/ds=\bm{d}$ represents the unit direction vector of the ray as shown in Fig.~\ref{fig-BOS-Schematic}. Integration of Eq.~(\ref{eq_RayEq}) over the ray path $S$ yields the light deflection after traversing PO as
\begin{equation}
\bm\varepsilon = n_{out}\bm{d}_{out}-n_{in}\bm{d}_{in}=\int_S \nabla n ds,
\label{eq_epsilonDefine_2}
\end{equation}
where $S$ is the optical path, $\bm\varepsilon$ is the deflection vector. It should be noted that the deflection vector $\bm\varepsilon$ is not the same as the deflection angle $\bm\theta$ as defined in Fig.~\ref{fig-BOS-Schematic}. For example, for a background plane perpendicular to the optical axis, $\bm\varepsilon =[\varepsilon_x, \varepsilon_y, \varepsilon_z]$ has three components, whereas $\bm\theta$ does not have $z$-component. 

Similar to the deflection estimation in 2D, a thin PO assumption gives,
\begin{equation}
\bm{d}_{in} \approx \hat{\bm{d}}_{in}.
%\label{eq:n0_bc}
\label{eq:din}
\end{equation}
For gaseous flow, further assume that
\begin{equation}
n_{in}=n_{out}=n_0.
\label{eq:n0_bc}
\end{equation}
Inside, $n_0$ denotes the refractive index at the volume boundaries. Consequently, the deflection vector estimation can be obtained as
\begin{equation}
\bm{\varepsilon} \approx n_0\bm{d}_{out}-n_0\hat{\bm{d}}_{in}=\int_S \nabla n ds.
\label{eq_epsilon_Tomoapprox}
\end{equation}
Discretization of Eq.~(\ref{eq_epsilon_Tomoapprox}) within the voxel-based framework gives,
\begin{equation}
\bm{\varepsilon}_{u}=S \cdot \bm{\nabla}_{u} \bm{n},
\label{eq_Deflection_discrete}
\end{equation}
where subindex $u\in\{x, y, z\}$ represents directional component. For $I$ rays pass through $J$ voxels in total, the array $\bm{\varepsilon}_{u} \in \mathbb{R}^{I}$ collects the $u$-direction deflection amounts of all rays from all views; the array $\bm{\nabla}_{u} \bm{n} \in \mathbb{R}^{J}$ is the discrete gradient in the $u$-direction; $S \in \mathbb{R}^{I \times J}$ is the tomographic projection matrix, in which the elements in the $i$-th row and $j$-th column represents the contribution of the refractive index gradient of the $j$-th voxel to the deflection of $i$-th ray. Therefore, Eq.~(\ref{eq_Deflection_discrete}) forms a linear system of equations in the form $ A\bm{x} = \bm{b}$. Equation~(\ref{eq_Deflection_discrete}) is usually ill-posed. Such an inverse problem is typically solved using regularization techniques to avoid amplifying noise. For the selection of specific tomographic algorithms and the design of regularization terms, see recent works \citep{grauer_fast_2020, liu_volumetric_2021, li_three-dimensional_2024}.

\subsection{Unified Deflection Estimation}\label{subsec: Deflection_estimation}

Comparing Eq.~(\ref{eq_epsilonDefine}) with Eq.~(\ref{eq_epsilon_Tomoapprox}), the identical right-hand sides of the equations indicate that the deflection vector component $\varepsilon_y$ is closely related to the angle $\theta_y$ extended by  $\bm{d}_{out}$ and $\hat{\bm{d}}_{in}$. To reveal this connection explicitly, based on Eq.~(\ref{eq_epsilon_Tomoapprox}), we write the $y$-component of the deflection vector $\varepsilon_y$ as,
\begin{equation}
{\varepsilon}_y \approx n_0\bm{j} \cdot (\bm{d}_{out} - \hat{\bm{d}}_{in})= \bm{j} \cdot \int_S \nabla n ds = \int_S \frac{\partial n}{\partial y} ds
\label{eq_epsilonMethod-2-component}
\end{equation}
Inside $\bm{j}$ represents the unit vector along the $y$-direction. Using the vector dot product formula:
\begin{equation}
\bm{j} \cdot \bm{d} = |{\bm{j}}| |{{\bm{d}}}| \cos\theta_{jd} = \cos\theta_{jd},
\label{eq: J_vectordot}
\end{equation}
where $\theta_{jd}$ is the angle between the directional vector $\bm{d}$ and vertical unit vector $\bm{j}$. Thus for 2D scenario, Eq.~(\ref{eq_epsilonMethod-2-component}) can be expressed in terms of $\theta$ as shown in Fig.~\ref{fig-BOS-Schematic},
\begin{equation}
\varepsilon_y \approx n_0\cos\theta_{out}-n_0\cos \hat{\theta}_{in},
\label{eq: n0cos}
\end{equation}
Compared to the definition of $\theta_y = \theta_{out}- \hat{\theta}_{in}$, it is clear that the deflection vector component $\varepsilon_y$ does not denote the physical angle $\theta_y$ as widely interpreted as in Fig.~\ref{fig-BOS-Schematic}.

By applying paraxial assumptions that rays are not deviated significantly from the optical axis, and assuming $n_0 \approx 1$, introducing $\beta_{out} =\frac{\pi}{2}-\theta_{out}$ and $\hat{\beta}_{in} =\frac{\pi}{2}-\hat{\theta}_{in}$, we can obtain:
\begin{equation}
\varepsilon_y \approx n_0\sin\beta_{out}-n_0\sin\hat{\beta}_{in}\approx  \beta_{out}-\hat{\beta}_{in} = \theta_y \approx \int_z \frac{\partial n}{\partial y} dz.
\label{eq: J_vectordot2}
\end{equation}
Equation (\ref{eq: J_vectordot2}) explains why most 2D BOS research treats $\varepsilon_y$ as an angle and estimates its value using the triangular relationship:
\begin{equation}
\varepsilon_y \approx \theta_y \approx \tan\theta_y \approx \Delta_y^\prime/Z_d.
\label{eq: J_vectordot3}
\end{equation}
Owing to the simplicity of Eq.~(\ref{eq: J_vectordot3}), such deflection estimations for 2D BOS are often directly extended to TBOS deflection estimations as well. 

\begin{figure}[htbp]
\centering
\includegraphics[width=0.5\textwidth]{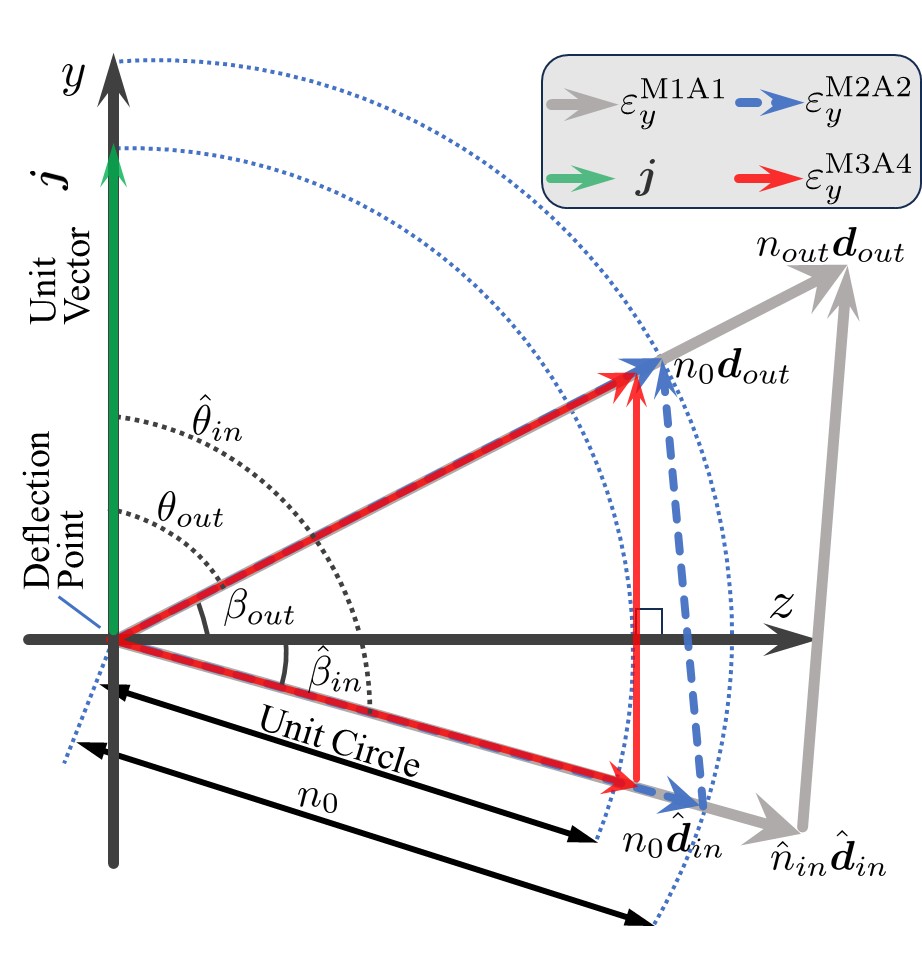}
\caption{{Geometric schematic for the three BOS deflection estimation methods in the unified framework in 2D scenario.}}\label{fig-DeflGeo}
\end{figure}

Clearly, various assumptions underlie the deflection estimation method in Eq.~(\ref{eq: J_vectordot3}). In the following, akin to the thin PO assumption, a 2D unified framework is firstly established with merely the variables $\beta_{out}$ and $\hat{\beta}_{in}$ as shown schematically in Fig.~\ref{fig-DeflGeo}. Different deflection estimation methods can be derived depending on the degree of simplification involved. 

The first estimation method can be derived as shown in Eq.~(\ref{eq: J_vectordot2}) as:
\begin{equation}
\varepsilon_{y}^\text{M1A1} = n_{out}\sin \beta_{out} - \hat{n}_{in}\sin \hat{\beta}_{in}.
\label{eq_unifiedFramework-M1}
\end{equation}
where $\hat{n}_{in}$ is at the position where the ray enters the PO, determined based on $\hat{\bm{d}}_{in}$. This derivation involves only one assumption that the thickness of PO is thin, thus termed as `M1A1' as marked in the superscript of $\varepsilon_{y}$ in Eq.~(\ref{eq_unifiedFramework-M1}). M1A1 is highly accurate in practice because most POs have negligible thickness relative to $Z_B$. Nevertheless, applying Eq.~(\ref{eq_unifiedFramework-M1}) is usually difficult for complex flows where $\hat{n}_{in}$ and $n_{out}$ are rarely precisely known at the boundaries. This also explains why most TBOS experiments, in order to apply M1A1 safely, require remaining constant peripheral atmospheric $n_0$ BCs, and also require removing light rays passing through the boundaries where $n$ is unknown \citep{nicolas_direct_2016}. 

The second deflection estimation method is a direct derivation from Eq.~(\ref{eq_unifiedFramework-M1}) as,
\begin{equation}
\varepsilon_{y}^\text{M2A2} = n_0\sin \beta_{out} - n_0\sin \hat{\beta}_{in}.
\label{eq_unifiedFramework-M2}
\end{equation}
with one additional assumption that $n_{in} = \hat{n}_{in}=n_{out} = n_0$. Thus, estimation method in Eq.~(\ref{eq_unifiedFramework-M2}) is termed as `M2A2'. For TBOS applications in complex flows, the peripheral $n$ BCs can deviate significantly from $n_0$. For example, in compressible flows, light rays often pass through the low-density side before the shock wave and leave the flow field from the higher-density side after the shock. In such cases, errors are expected with the M2A2 method, and careful error analysis is required.

The last deflection estimation method can be derived from Eq.~(\ref{eq: J_vectordot3}) using $\beta_{out}$ and $\hat\beta_{in}$ according to the schematic in Fig.~\ref{fig-BOS-Schematic} as:
\begin{equation}
%\varepsilon_{y}^{M3A4} = \frac{\Delta'_y}{Z_{d}} = \frac{B^\prime O}{Z_{d}} + \frac{OB}{Z_{d}} = \tan \beta_{out} - \tan \beta_{in}
\varepsilon_{y}^\text{M3A4} = \frac{\Delta'_y}{Z_{d}} = \frac{BO}{Z_{d}} - \frac{B^\prime O}{Z_{d}} = \tan\beta_{out} - \tan\hat{\beta}_{in}.
\label{eq_unifiedFramework-M3}
\end{equation}
This estimation method is termed M3A4 because it involves two additional assumptions: the paraxial assumption and the perpendicularity condition between the deflection vector and the principal optical axis. Consequently, the error of applying M3A4 can be amplified when the principal optical axes of the cameras are non-perpendicular to the background plane, widely encountered for TBOS applications with multiple cameras but only a few background planes \citep{nicolas_experimental_2017, grauer_instantaneous_2018, weisberger_preparations_2020, bo_background-oriented_2023}. 

In summary, four assumptions are involved in the three deflection estimation methods for BOS as in Table~\ref{table_unified_table}. First, a thin PO or a fixed deflection plane; Second, $n_0$ at boundaries; Third, the paraxial assumption; Fourth, the perpendicularity condition between the deflection vector and the principal optical axis. Moreover, Eqs.~(\ref{eq_unifiedFramework-M1})--(\ref{eq_unifiedFramework-M3}) unify the typical deflection estimation methods for both 2D-BOS using only $\beta_{out}$ and $\hat{\beta}_{in}$ in the trigonometric forms and associated schematic is shown in Fig.~\ref{fig-DeflGeo}.

\begin{table}[!htb]
\caption{Assumptions involved within the three deflection estimation methods for 2D-BOS (Take $\varepsilon_y$ as an example for the triangular form).}
\label{table_unified_table}
\centering
\begin{tabular}{cccc}
\hline 
\text{Methods} & \text{Assumptions} & \text{Vector Form} & \text{Triangular Form} \\
\hline 
\text{Theoretical} & \text{None} & $n_{\text{out}} \bm{d}_{\text{out}} - n_{\text{in}} \bm{d}_{\text{in}}$ & $n_{\text{out}} \sin \beta_{\text{out}} - n_{\text{in}} \sin \beta_{\text{in}}$ \\
\text{M1A1} & 1 & $n_{\text{out}} \bm{d}_{\text{out}} - \hat{n}_{\text{in}} \hat{\bm{d}}_{\text{in}}$ & $n_{\text{out}} \sin \beta_{\text{out}} - \hat{n}_{\text{in}} \sin \hat{\beta}_{\text{in}}$ \\
\text{M2A2} & 1,2 & $n_0 \bm{d}_{\text{out}} - n_0 \hat{\bm{d}}_{\text{in}}$ & $n_0 \sin \beta_{\text{out}} - n_0 \sin \hat{\beta}_{\text{in}}$ \\
\text{M3A4} & 1,2,3,4 & $\left[\Delta_x^{\prime} / Z_d, \Delta_y^{\prime} / Z_d \right]$ & $\tan \beta_{\text{out}} - \tan \hat{\beta}_{\text{in}}$ \\
\hline
\end{tabular}
\end{table}

The complete unified framework for 3D arbitrary deflection vectors are also established and derived in the Appendix \ref{sec_A1}. 

\section{Nonlinear Ray Tracing Platform}\label{sec:RayTracing}

To quantitatively analyze errors arising from ray deflection in BOS experiments, a high-fidelity nonlinear ray-tracing simulation (RTS) platform, adapted from \citep{rajendran_pivbos_2019}, is used \citep{li_three-dimensional_2024, li_neural_2024, jia_tomographic_2024}. The specific modules include light-field generation, nonlinear light propagation through the PO, light deflection by the imaging optical elements in accordance with Snell's law, and image rendering based on a light-diffraction model. Detailed implementations can be found in the literature cited above and are not repeated here. Nevertheless, to enable evaluation of the BOS deflection estimation error, rather than rendering synthetic BOS images, ray trajectories and intersection points on the sensor plane are recorded directly to compute the deflection and displacement vectors. Two types of POs are designed for the RTS as detailed below.

\subsection{Chirp-type Phase Object} \label{subsec_Chirp-PO}

\begin{figure}[htbp]
\centering
\includegraphics[width=0.5\textwidth]{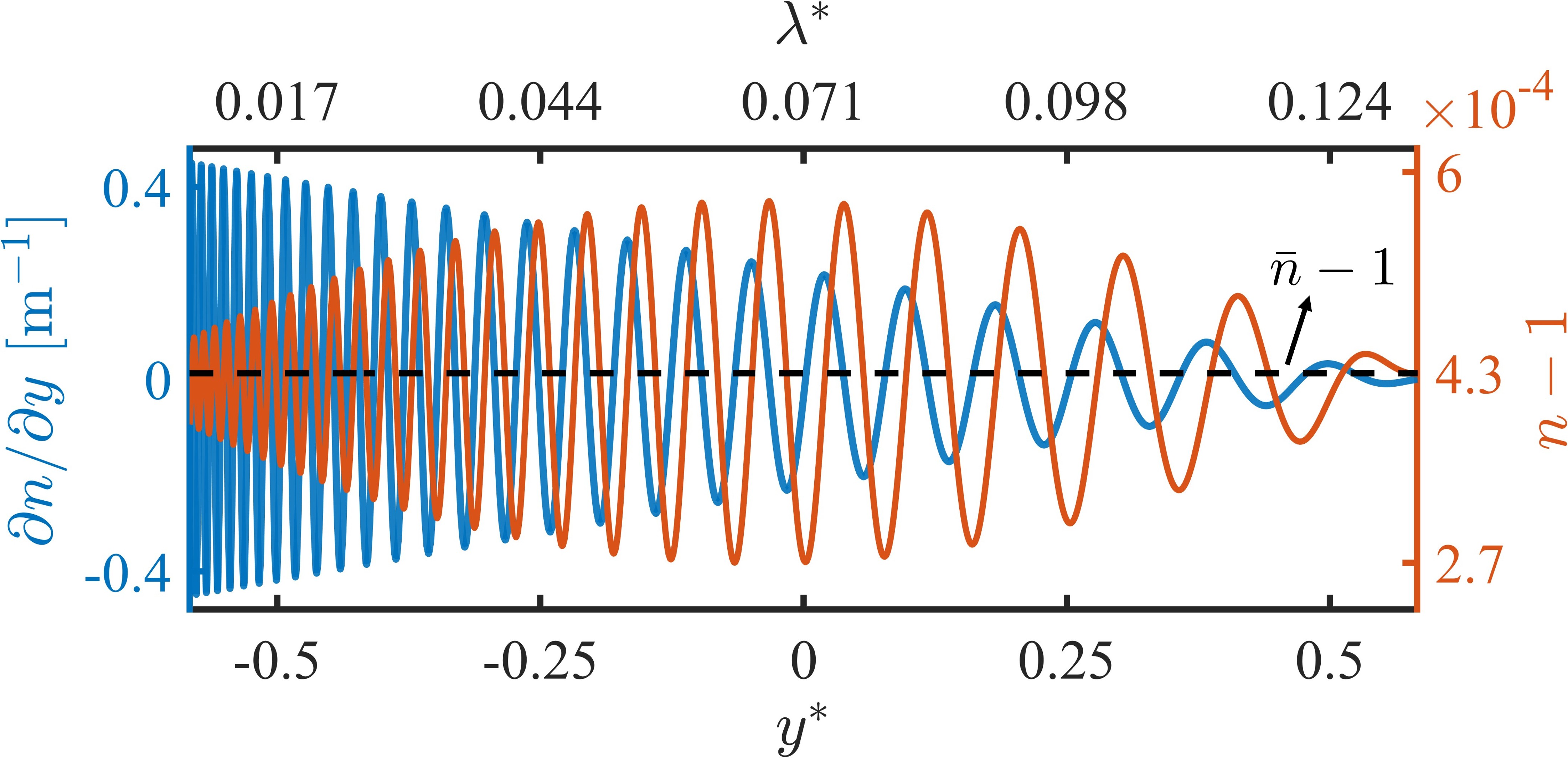}
\caption{Chirp-type $\partial n / \partial y$ and $ n $ generated according to Eqs.~(\ref{eq_ChirpDefinition})--(\ref{eq_ChirpDefinition-Lambda}).}\label{fig-ChirpSignal}
\end{figure}

Consider a one-dimensional chirp $\partial n/\partial y$ distribution \citep{zhou_comparison_2025} of linearly decreasing amplitude $A$ and wavelength $\lambda$ according to the equations,
\begin{equation}
\frac{\partial n}{\partial y}(y) = A(y+\frac{L_c}{2}) \sin \phi(y+\frac{L_c}{2}),
\label{eq_ChirpDefinition}
\end{equation}
\begin{equation}
A(y) = G_m \frac{L_c-y}{L_c},
\label{eq_ChirpDefinition-Adecrease}
\end{equation}
\begin{equation}
\phi(y)=\int_0^y \frac{2 \pi}{\lambda(x)} d x,
\label{eq_ChirpDefinition-phi}
\end{equation}
\begin{equation}
\lambda(y) = \lambda_{min} + \frac{\lambda_{max} - \lambda_{min}}{L_c}(y+\frac{L_c}{2}).
\label{eq_ChirpDefinition-Lambda}
\end{equation}
Inside, $G_m = \max({\partial n}/{\partial y})= 0.45 ~\text{m}^{-1}$, $L_c = 70$ mm is the total length of the chirp signal; $\lambda_{max} = 1/125 ~\text{m}$ and $\lambda_{min} = 1/2000 ~\text{m}$, $\phi$ is the phase of the waveform. The resulting ${\partial n}/{\partial y}$ and $n$ distributions are shown in Fig.~\ref{fig-ChirpSignal}. Note that $n$ is obtained by integrating ${\partial n}/{\partial y}$, and its minimum value is set to $n_0 = 1.00027$. Normalized quantities are defined as $\lambda^* = \lambda/W_F$, and $y^* = y/W_F$ where $W_F$ is the field of view width at $z=0$ mm and is set to 60 mm. 

\begin{figure}[htbp]
\centering
\includegraphics[width=1\textwidth]{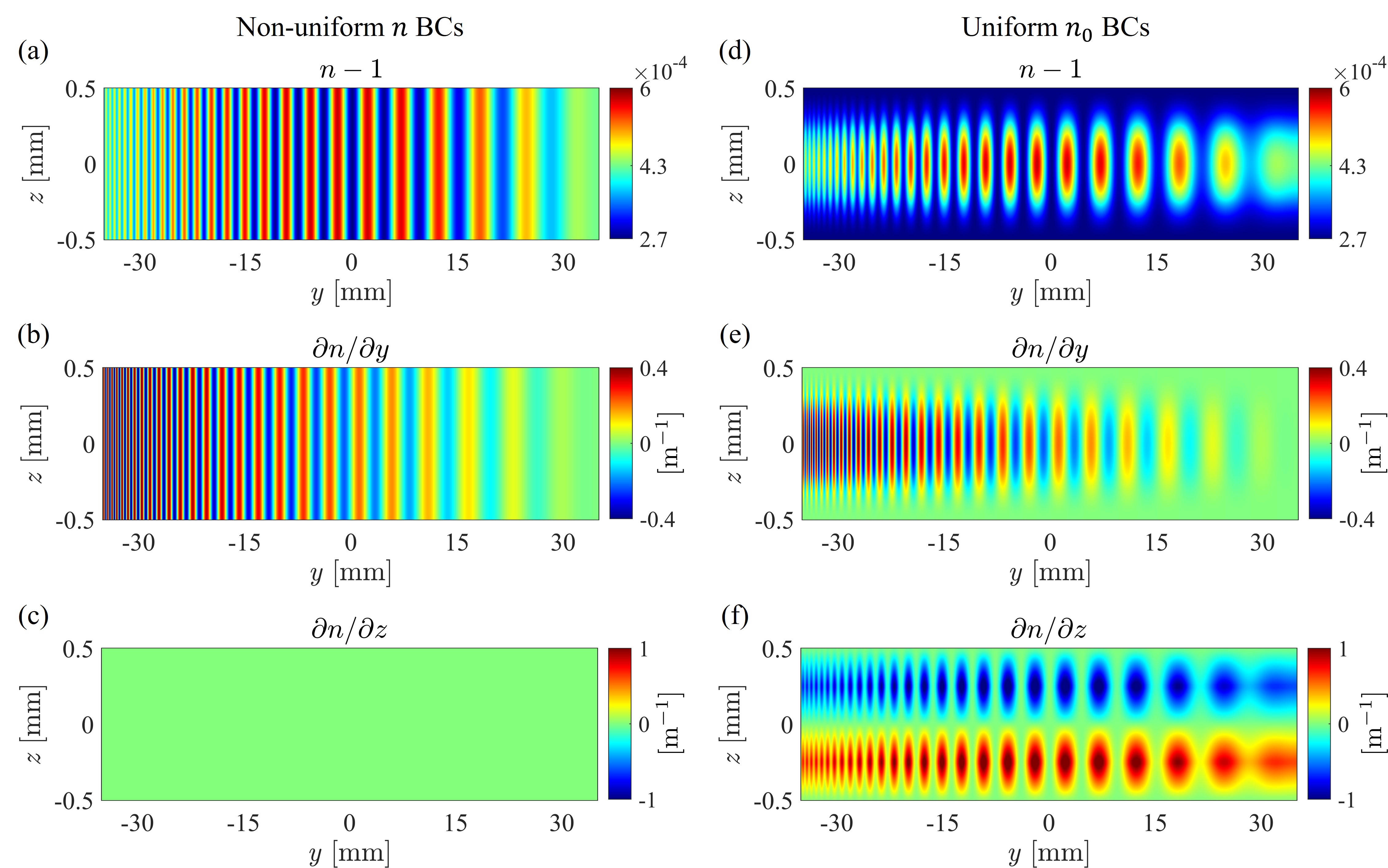}
\caption{Chirp distributions of $n$, $\partial n/\partial y$,  $\partial n/\partial z$ with (a)--(c) Non-uniform $n$ BCs, and (d)--(f) Uniform $n_0$ BCs.}
\label{fig-Chirp-Type-PO}
\end{figure}

To construct chirp-type POs of finite thicknesses with different boundary conditions (BCs), two strategies for $n$ slice extension along the optical axis are considered. The first BC type features a constant extension of the chirp-type $n$ slice in Fig.~\ref{fig-Chirp-Type-PO}(a)--(c) in the z-direction. Consequently, non-uniform $n$ is expected at the front and back faces of the slab at $z=\pm0.5$ mm. Thus, this case involves non-uniform $n$ BCs. The second type has a quasi-Gaussian extension of $n$ in the $z$-direction, arriving at a constant $n_0$ at $z=\pm0.5$ mm; thus, uniform $n_0$ BCs are achieved. The two PO slabs above are designed to reveal the impact of the $n$ BCs on deflection estimation.

\subsection{Turbulent-type Phase Object} \label{subsubsec_Turbulent-PO}

\begin{figure}[htbp]
\centering
\includegraphics[width=1\textwidth]{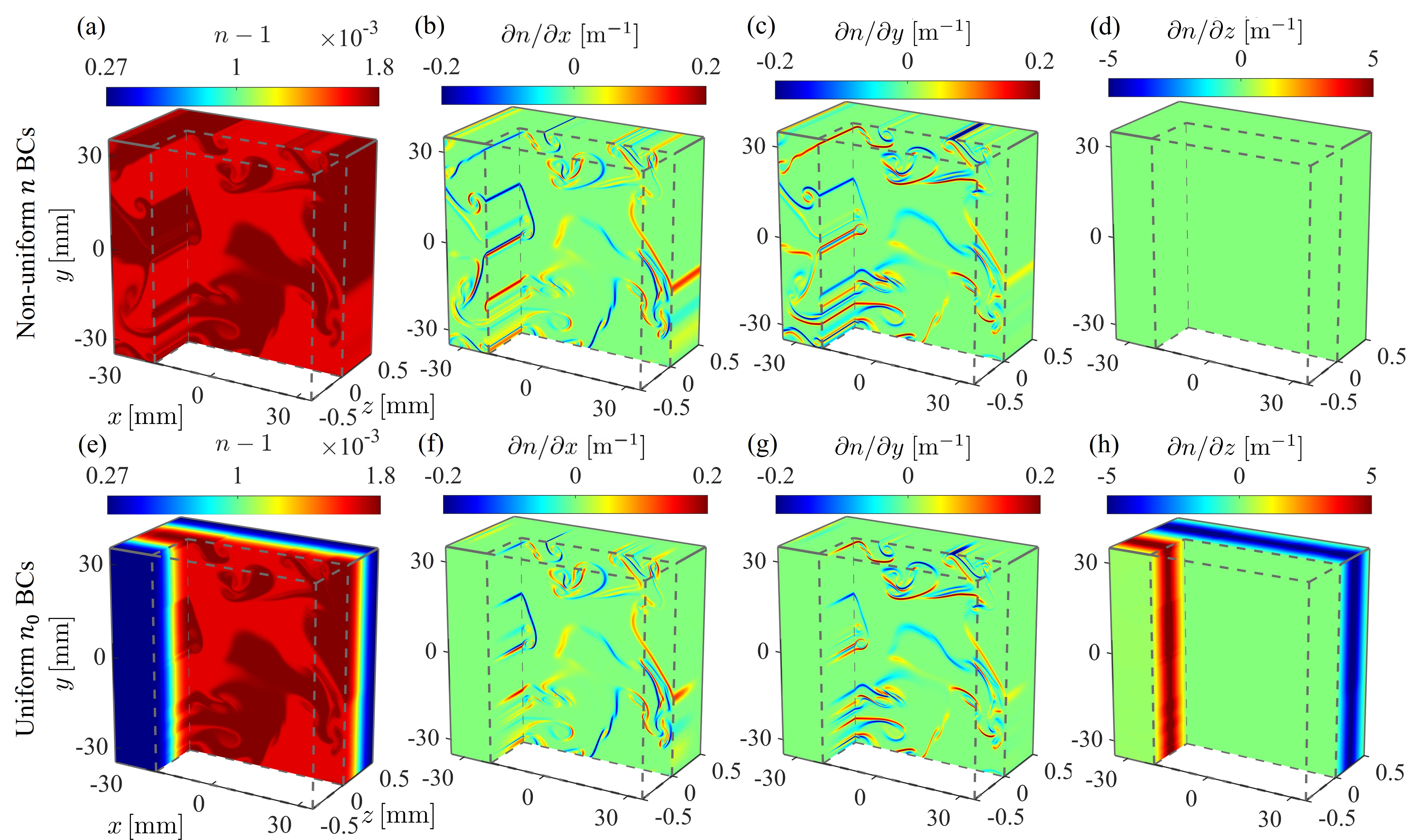}
\caption{Turbulent-type POs with (a)--(d) non-uniform $n$ BCs and (e)--(h) uniform $n_0$ BCs.}\label{fig-turbulent-type-PO}
\end{figure}

To assess deflection estimation in more realistic flow scenarios, an instantaneous slice of the buoyancy-driven turbulence from the Johns Hopkins University direct numerical simulation dataset \citep{li_public_2008} is used to design the turbulent-type POs. This flow type is selected owing to the complex multi-scale $\rho$ structure. Similar to the design of chirp-type POs, the turbulent slice is also extended either constantly or with quasi-Gaussian distributions along the z-direction to enable uniform and non-uniform $n$ BCs, as shown in Fig.~\ref{fig-turbulent-type-PO}. 

\subsection{Deflection and Displacement Extraction} \label{subsec_def_disp}

During the nonlinear RTS, the above-designed POs are allocated midway between the sensor plane and the background to maximize the measurement sensitivity ($Z_d = Z_a = 0.5Z_B$) \citep{lang_measurement_2017}. Note that for the chirp-type POs, ray deflections occur only in the $y$- and $z$-directions, while for the turbulent-type ones, ray deflections occur in all directions. To minimize the influence of the non-linear ray trajectories, all POs are designed with a thin thickness $L = 1$ mm ($L/Z_B \approx 5.6 \times 10^{-4}$). Detailed settings for the BOS nonlinear RTS are provided in Table~\ref{table_BOSconfig}, where $f$ represents the imaging lens focal length and $W_S$ denotes the image sensor size.

\begin{table}[!htb]
\caption{Parametric settings for BOS nonlinear RTS.}
\label{table_BOSconfig}
%\begin{tabular}{@{}lll@{}}
\begin{tabular}{@{}lcc@{}}
\toprule
\multirow{2}{*}{Parameter} & \multicolumn{2}{c}{Setting} \\
\cmidrule(l){2-3}
& Chirp-type POs & Turbulent-type POs \\
\midrule
Imaging model & \multicolumn{2}{c}{Pinhole} \\
Ray tracing algorithm & \multicolumn{2}{c}{$4^{\text{th}}$ Runge-Kutta method}\\
Ray tracing step size & \multicolumn{2}{c}{$4 \times 10^{-4}$ mm/step}\\
$Z_{B}$ & \multicolumn{2}{c}{1800 mm} \\
$Z_{d}$ & \multicolumn{2}{c}{900 mm} \\
$f$ & \multicolumn{2}{c}{50 mm} \\
$Z_{i}$ & \multicolumn{2}{c}{51.43 mm} \\
$W_{S}$ & ($y$) 3.43 mm & ($x \times y$) $3.43 \times 3.43$ mm\\
$W_{F}$ & ($y$) 60 mm & ($x \times y$) $60 \times 60$ mm \\
PO Size & ($y \times z$) $70 \times 1$ $\mathrm{mm^2}$ & ($x \times y \times z$) $70 \times 70 \times 1$ $\mathrm{mm^3}$ \\
PO Resolution & ($y \times z$) $35000 \times 50$ voxels & ($x \times y \times z$) $500 \times 500 \times 50$ voxels \\
\botrule
\end{tabular}
\end{table}

\begin{figure}[htb]
\centering
\includegraphics[width=0.5\textwidth]{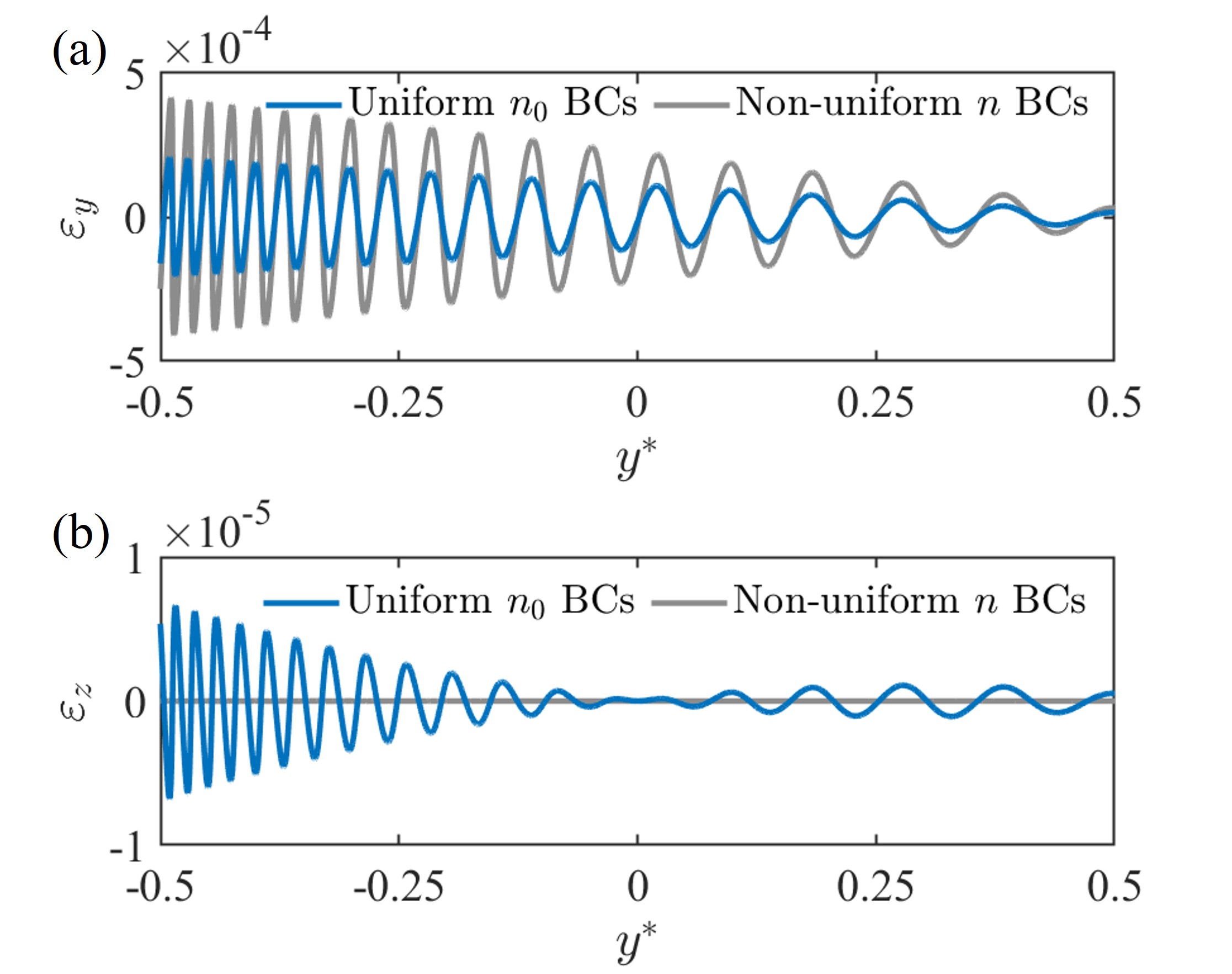}
\caption{Deflection vector (a)$\varepsilon_y$ and (b)$\varepsilon_z$ for chirp-type POs with different $n$ BCs. }\label{fig-Defl_RT_chirp}
\end{figure}

\begin{figure}[htb]
\centering
\includegraphics[width=1\textwidth]{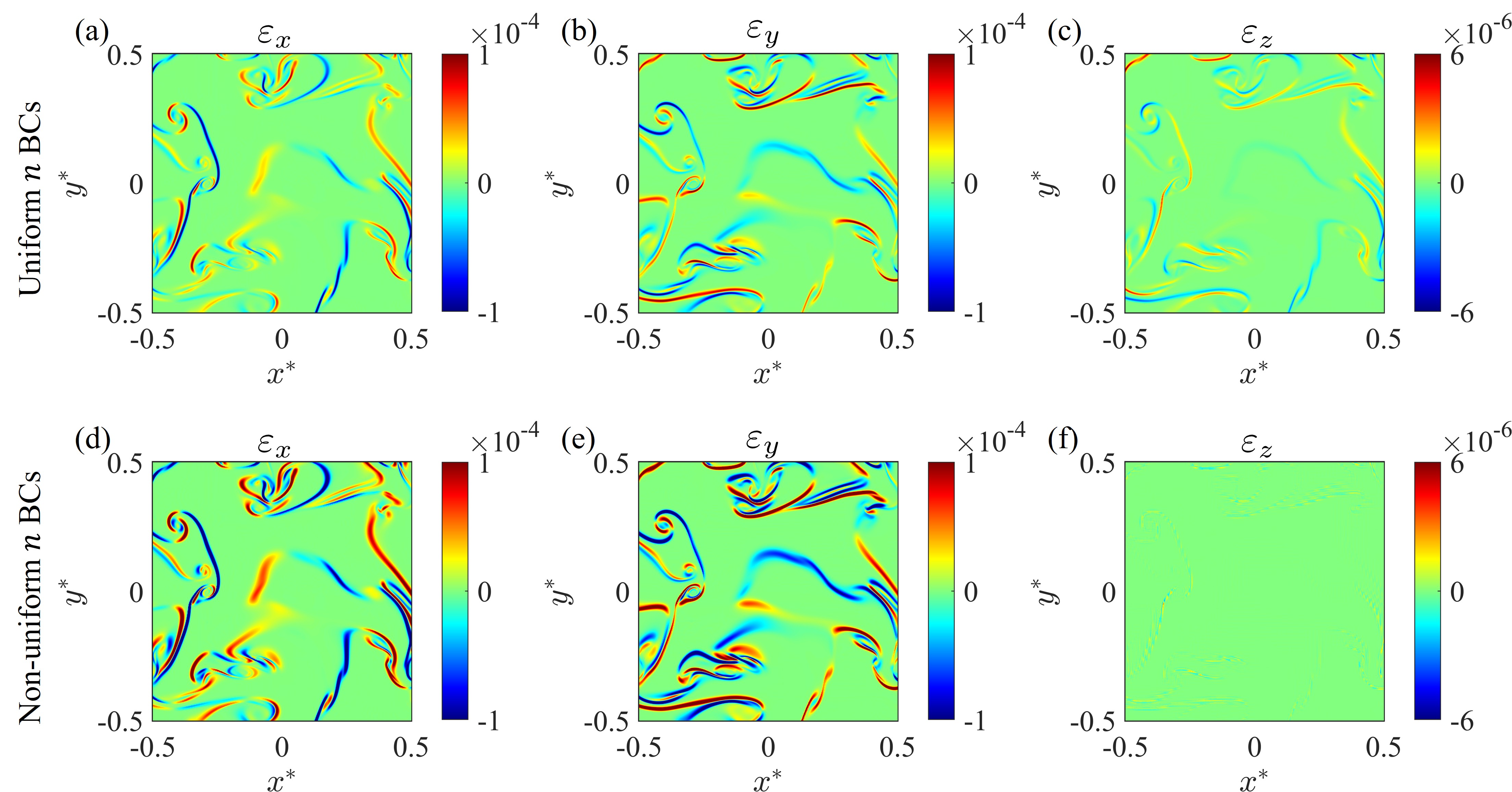}
\caption{Deflection vector $\varepsilon_x$,$\varepsilon_y$ and $\varepsilon_z$ for turbulent-type POs with  (a)--(c) uniform $n_0$ and (d)--(f) non-uniform $n$  BCs.}\label{fig-Defl_RT_JHU}
\end{figure}

Owing to the RTS, ray trajectories within the POs can be recorded, and resulting deflection vectors can be computed straightforwardly from Eq.~(\ref{eq_epsilonDefine_2}) as illustrated in Figs.~\ref{fig-Defl_RT_chirp} and \ref{fig-Defl_RT_JHU} for chirp-/turbulent-type POs, respectively. Obviously, for the chirp-/turbulent-type POs, ray deflection occurs in two/three-directions, respectively. Furthermore, for the POs with uniform $n_0$ BCs, a non-zero $\varepsilon_z$ can be expected as shown in Fig.~\ref{fig-Defl_RT_chirp}(b) and \ref{fig-Defl_RT_JHU}(c). As a contrast, for POs with non-uniform $n$ BCs, $\varepsilon_z = 0$ holds as confirmed in Figs.~\ref{fig-Defl_RT_chirp}(b) and \ref{fig-Defl_RT_JHU}(f) due to the uniform extension of $n$ slices in the $z$-direction. By further tracking intersection positions between the outgoing rays and the sensor with and without POs, the exact displacement $\Delta(x,y)$ can be determined. The spatial distribution of $\Delta(x,y)$ is similar to those shown in Figs.~\ref{fig-Defl_RT_chirp} and \ref{fig-Defl_RT_JHU}, thus not duplicated here. Deflection vectors and displacement fields obtained by RTS serve both as the reference ground truth and as input to the deflection estimation methods as detailed later.

\section{Error Analysis on Deflection Estimation}\label{sec_Results}

After describing the three BOS deflection estimation methods within the unified framework, we proceed to a detailed error analysis of each method. Two error indices, the absolute error and the relative error, are defined as
$\epsilon_{\tiny{(x,y,z)}}^\text{MaAb} =\varepsilon_{(x,y,z)}^\text{RTS} - \varepsilon_{(x,y,z)}^\text{MaAb}$ and the relative one $\hat{\epsilon}_{(x,y,z)}^\text{MaAb}=\epsilon_{(x,y,z)}^\text{MaAb}/\varepsilon_{(x,y,z)}^\text{RTS}$, respectively. Inside the superscripts $a = {1,2,3}$ and $b = 1,2,4$ to denote the three estimation methods (details in Table~\ref{table_unified_table}); the subscripts $(x, y, z)$ denote the deflection components in the three directions. Note that for the chirp-type POs, only the $y$-/$z$-component exists as claimed in Sec.~\ref{subsec_def_disp}. For turbulent-type POs, many regions are of $\varepsilon_{(x,y,z)}^\text{RTS} = 0$ leading to division-by-zero issues in the calculation $\hat{\epsilon}_{z}$, thus not explored. For chirp-type POs, low-pass filtering and thresholding are used to remove these division-by-zero-induced spikes in $\hat{\epsilon}_{z}$.
%%%%%%%%%%%%%%%%%%%%%%%%%%%%%%%%%%%%%%%%%%%%%%%%%%%%%

\subsection{M1A1 Method}\label{sec_M1A1}

\begin{figure}[htb]
\centering
\includegraphics[width=1\textwidth]{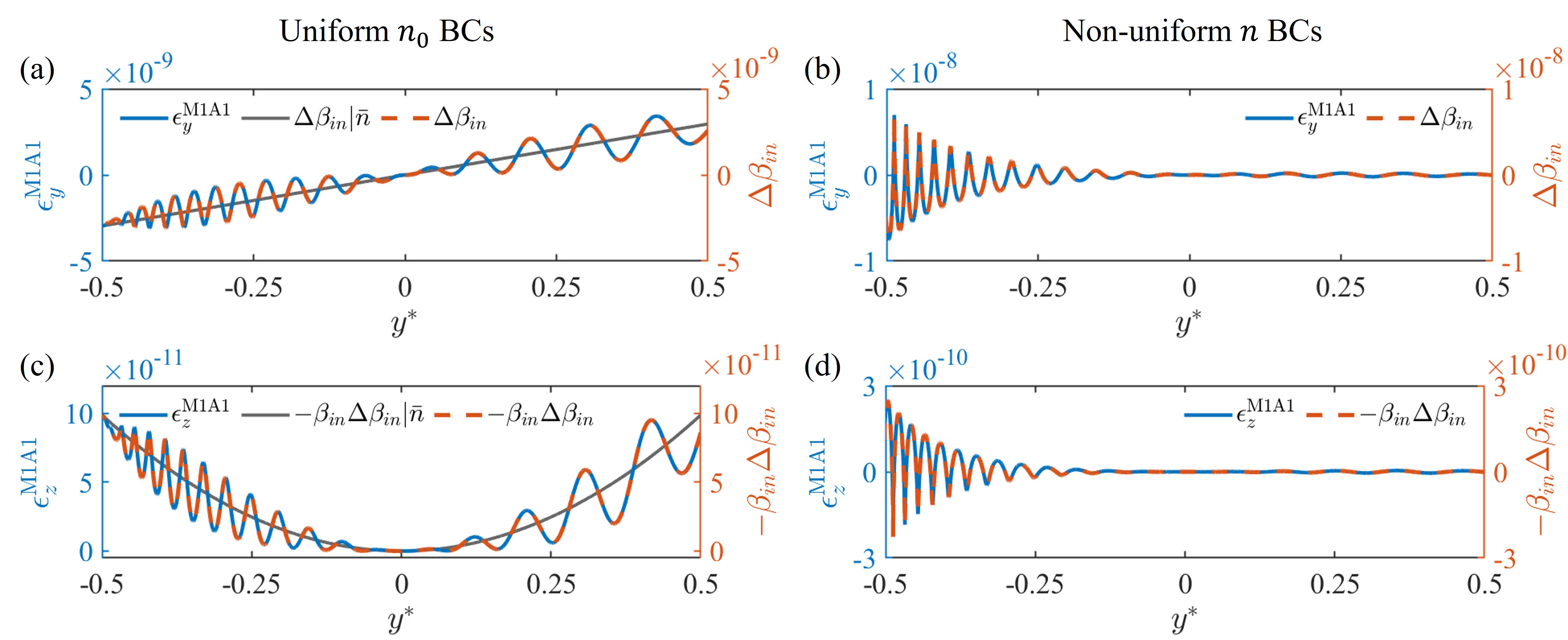}
\caption{$\epsilon_{(y,z)}^\text{M1A1}$ for the chirp-type POs with both uniform/non-uniform $n$ BCs.}
\label{fig-DeflError_RT-M1A1_chirp}
\end{figure}

Regarding the errors associated with the M1A1 method, the PO thickness $L$ is set only to be 1 mm ($L/Z_B \approx 5.6 \times 10^{-4}$), making the thin PO assumption hold well. To confirm the superiority of the M1A1 method, $\epsilon_{(y,z)}^\text{M1A1}$ are plotted in Fig.~\ref{fig-DeflError_RT-M1A1_chirp} for the chirp-type POs. Overall, two observations hold regardless of the BCs. First, ultra-low level of $\epsilon_{(y,z)}^\text{M1A1}$ can be identified compared to the ones predicted by M2A2 and M3A4 methods as detailed later; Second, the spatial frequencies of $\epsilon_{(y,z)}^\text{M1A1}$ resemble those of $\nabla n$ for the chirp-type POs. 

Moreover, unique features are akin to the behaviors of $\epsilon_{(y,z)}^\text{M1A1}$ with each type of BCs. For uniform $n_0$ BCs, a global linear/parabolic trend can be identified in $\epsilon_{y}^\text{M1A1}$/$\epsilon_{z}^\text{M1A1}$ respectively, as shown in Fig. \ref{fig-DeflError_RT-M1A1_chirp}. To reveal the mechanism behind, the triangular form of $\epsilon_{(y,z)}^\text{M1A1}$ with uniform $n_0$ BCs is firstly derived according to the definition under the unified framework as,
\begin{equation}
\epsilon_{y}^\text{M1A1} = n_0 (\sin \beta_{out} - \sin \beta_{in}) - n_0 (\sin \beta_{out} - \sin \hat{\beta}_{in}) = n_0 (\sin \hat{\beta}_{in} - \sin \beta_{in})
\label{eq_DeflError_RT-M1_UBC_y1}
\end{equation}
Defining the incident angle difference $\Delta \beta_{in} = \hat{\beta}_{in} - \beta_{in}$, which is assumed to be a small quantity for a thin phase object, and applying the Taylor expansion to Eq.~(\ref{eq_DeflError_RT-M1_UBC_y1}), we can obtain,
\begin{equation}
\epsilon_y^\text{M1A1} = n_0 [\sin(\beta_{in} + \Delta \beta_{in}) - \sin \beta_{in}] \approx n_0 \cos\beta_{in} \Delta \beta_{in}
\label{eq_DeflError_RT-M1_UBC_y2}
\end{equation}
Similarly, the triangular form of $\epsilon_z^\text{M1A1}$ yields,
\begin{equation}
\epsilon_z^\text{M1A1} = n_0 ( \cos \hat{\beta}_{in} - \cos \beta_{in}) \approx -n_0 \sin\beta_{in} \Delta \beta_{in}
\label{eq_DeflError_RT-M1_UBC_z}
\end{equation}
Thus, the contrasting linear/parabolic global trends of $\epsilon_{(y,z)}^\text{M1A1}$ can be explained by the $\cos$/$\sin$ functions within the Eq.(\ref{eq_DeflError_RT-M1_UBC_y2}) and Eq.(\ref{eq_DeflError_RT-M1_UBC_z}). Moreover, since $\beta_{in}$ is a small quantity, $\cos\beta_{in}\approx1$ and $\sin\beta_{in}\approx \beta_{in}$ thus $\epsilon_y^\text{M1A1} \approx \Delta \beta_{in}$ and $\epsilon_z^\text{M1A1} \approx -\beta_{in}\Delta \beta_{in}$ as confirmed by Fig.\ref{fig-DeflError_RT-M1A1_chirp}(a) and (c).

To reveal the mechanism behind the global linear/parabolic trend of $\epsilon_{(y,z)}^\text{M1A1}$, we construct a special $n$ field featured by the mean $n$ value of the chirp field, $\bar{n}$, as indicated by the black dash line in Fig.~\ref{fig-ChirpSignal}. By replacing the chirp $n$ field by $\bar{n}$ as the peak of the Gaussian distribution with the uniform $n_0$ BC. Resulting $\epsilon_{(y,z)}^\text{M1A1}$ exhibits exactly the linear/parabolic trend as shown in Fig.\ref{fig-DeflError_RT-M1A1_chirp}(a) and (c) denoted by $\Delta\beta|\bar{n}$. This observation reveals that the global trend of $\epsilon_{(y,z)}^\text{M1A1}$ is controlled by the $\bar{n}$ and the specific distribution of $n$ along the optical axis. Consider that uniform $n_0$ BCs are always preferred within BOS experiments, and a quasi-Gaussian distribution along the optical axis is common, so global linear/parabolic errors can always be expected for $\epsilon_{(y,z)}^\text{M1A1}$.

Regarding the chirp-type POs with non-uniform $n$ BCs, firstly, we can observe that the global linear/parabolic trend of $\epsilon_{(y,z)}^\text{M1A1}$ vanishes as shown in Fig.\ref{fig-DeflError_RT-M1A1_chirp}(b) and (d). This confirms that it is mainly the Gaussian distribution of $n$ along the optical axis, not the off-principal-axis related error, responsible for the global linear/parabolic trend of $\epsilon_{(y,z)}^\text{M1A1}$. Moreover, much larger oscillating amplitude of $\epsilon_{(y,z)}^\text{M1A1}$, compared to that of the uniform $n_0$ BCs, indicates that non-uniform $n$ BCs introduce larger deflection estimation errors, even provided precise $n$ at BCs. To gain further insight, similar derivations of $\epsilon_{y}^\text{M1A1}$ with non-uniform $n$ BCs can be made within the unified framework as,
\begin{equation}
\epsilon_y^\text{M1A1} = n_{in}(\frac{\hat{n}_{in}}{n_{in}} \sin \hat{\beta}_{in} - \sin \beta_{in})
\label{eq_DeflError_RT-M1_NUBC_y1}
\end{equation}
For aerodynamic applications, $n_{in} \approx n_0\approx n_{in} / \hat{n}_{in} \approx 1$ and Eq.~(\ref{eq_DeflError_RT-M1_NUBC_y1}) can be further simplified as,
\begin{equation}
\epsilon_y^\text{M1A1} \approx n_{in}[ \sin(\beta_{in} + \Delta \beta_{in}) - \sin \beta_{in}] \approx n_{in} \cos\beta_{in} \Delta \beta_{in}
\label{eq_DeflError_RT-M1_NUBC_y2}
\end{equation}
Again, for small $\beta_{in}$, we can have $\epsilon_y^\text{M1A1} \approx \Delta \beta_{in}$ as confirmed in Fig. ~\ref{fig-DeflError_RT-M1A1_chirp}(b). Compared to $\Delta \beta_{in}$ with uniform BCs shown in Fig. ~\ref{fig-DeflError_RT-M1A1_chirp}(a), very different oscillating amplitudes and global trends can be identified, indicating the high sensitivity of $\Delta \beta_{in}$ over the $n$ BCs and the $n$ distributions along the optical axis. Similarly, the triangular form of $\epsilon_{z}^\text{M1A1}$ can also be derived as,
\begin{equation}
\epsilon_z^\text{M1A1} \approx -n_{in} \sin\beta_{in} \Delta\beta_{in}
\label{eq_DeflError_RT-M1_NUBC_z}
\end{equation}
which can be simplified as $\epsilon_z^\text{M1A1} \approx -\beta_{in}\Delta \beta_{in}$ as confirmed in Fig. ~\ref{fig-DeflError_RT-M1A1_chirp}(d).

\begin{figure}[htb!]
\centering
\includegraphics[width=1\textwidth]{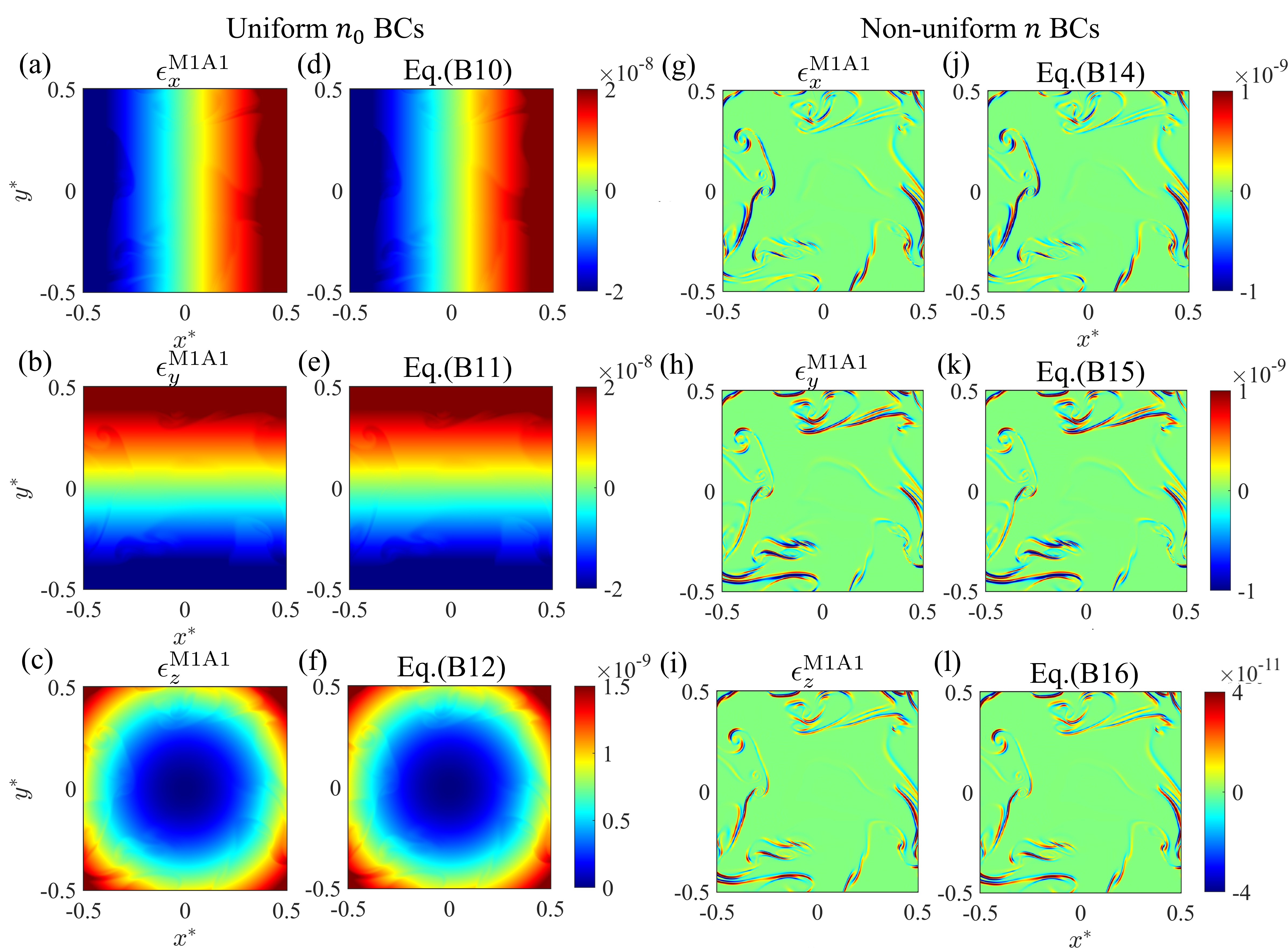}
\caption{Comparisons on $\epsilon_{(x,y,z)}^\text{M1A1}$ obtained from RTS(a-c)/(g-i) and derived triangular forms under the unified framework(d-f)/(j-l) for the turbulent-type POs with both uniform/non-uniform $n$ BCs.}\label{fig-Defl_RT-M1_JHU_epsilon}
\end{figure}

To explore the ${\epsilon}_{(x,y,z)}^\text{M1A1}$ for more realistic flows, turbulent-type POs are explored and RTS results of ${\epsilon}_{(x,y,z)}^\text{M1A1}$ are shown in Fig.~\ref{fig-Defl_RT-M1_JHU_epsilon}(a)--(c) and (g)--(i) for both uniform and non-uniform BCs. Related analytical triangular forms for $\epsilon_{(x,y,z)}^\text{M1A1}$ are derived within the unified framework in detail in Appendex.\ref{sec_A2} and predicted fields are also presented as in Fig.~\ref{fig-Defl_RT-M1_JHU_epsilon}(d)--(f) and (j)--(l). The near-perfect match between the RTS results and the analytical derivations confirms the applicability of the unified deflection estimation framework to multiscale density-varying flows. Similarly, with uniform $n_0$ BCs, global linear/parabolic trends can be again identified within the spatial distribution of ${\epsilon}_{(x,y,z)}^\text{M1A1}$, verifying the essential role of the Gaussian distribution of $n$ along the optical axis.

In sum, for the M1A1 method, ${\epsilon}_{(x,y,z)}^\text{M1A1}$ is mainly caused by the nonlinear ray trajectories within the POs. Provided with uniform $n_0$ BCs, the global trend of M1A1 errors exhibits a global linear/parabolic trend due to the Gaussian profile of $n$ along the optical axis. With non-uniform $n_0$ BCs, the amplitude of ${\epsilon}_{(x,y,z)}^\text{M1A1}$ boosts due to the fluctuation $n$ BCs. Nevertheless, the amplitude of ${\epsilon}_{(x,y,z)}^\text{M1A1}$ is still trivial with the current simulation setup. Without mentioning, we assume the deflection estimation based on the M1A1 method can replace the ground truth deflection, enabling the derivation of the analytical expression of the relative error $\hat{\epsilon}_{(x,y,z)}$ in the following.

%%%%%%%%%%%%%%%%%%%%%%%%%%%%%%%%%%%%%%%%%%%%%%%%%%%%%
\subsection{M2A2 Method}
Regarding the errors associated with the M2A2 method, first consider the POs with uniform $n_0$ BCs as shown in Fig.~\ref{fig-Chirp-Type-PO}(d)--(f) and Fig.~\ref{fig-turbulent-type-PO}(e)--(h). Since $n_{in} = n_{out} = n_0$, the M2A2 method now retreats to the M1A1 one so that $\epsilon_{(y,z)}^\text{M2A2} = \epsilon_{(y,z)}^\text{M1A1}$ holds. In the following, we only explore POs with non-uniform $n$ BCs as shown in Fig.~\ref{fig-Chirp-Type-PO}(a)--(c) and Fig.~\ref{fig-turbulent-type-PO}(a)--(c). 

\begin{figure}[htb]
\centering
\includegraphics[width=0.55\textwidth]{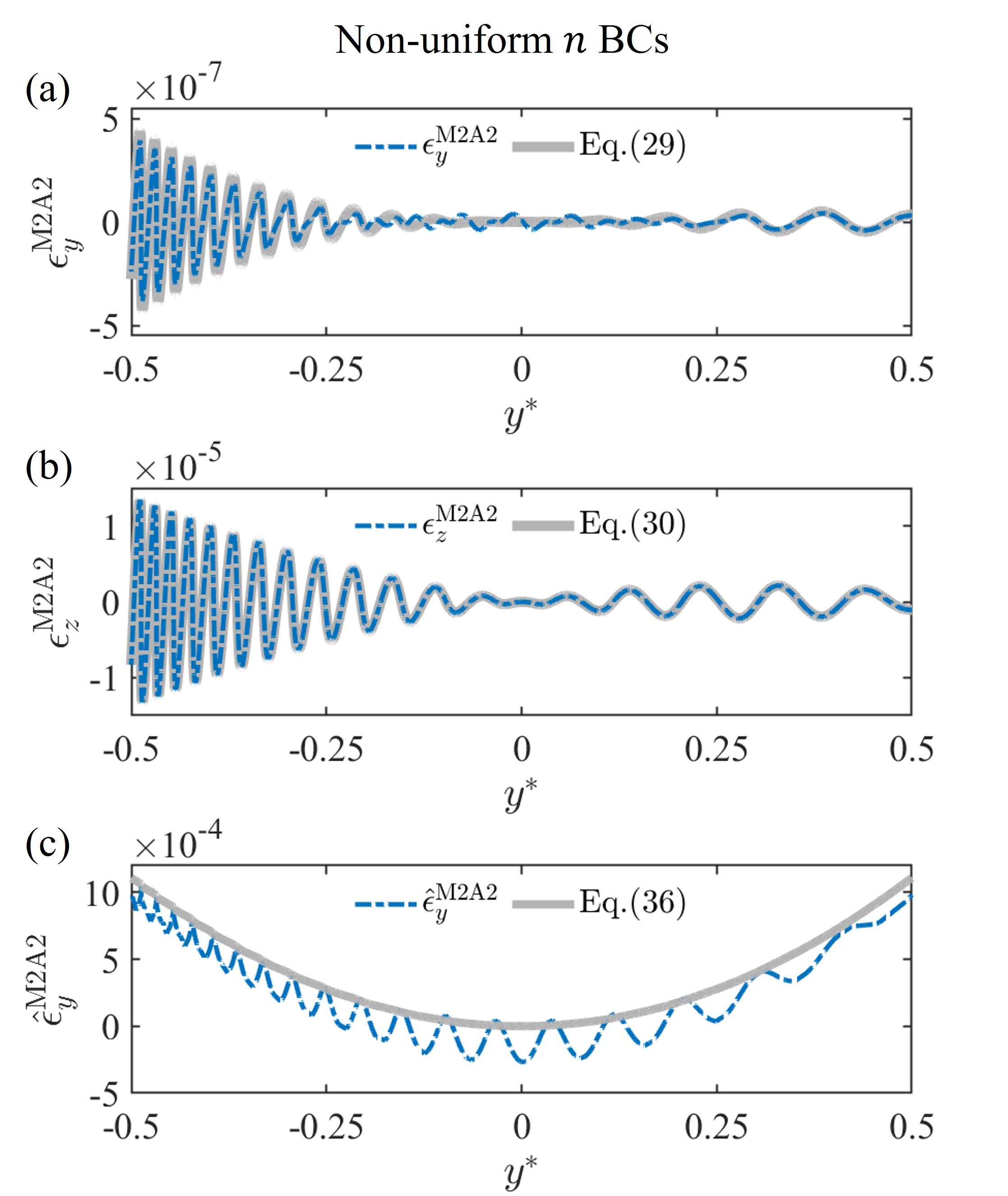}
\caption{Comparisons on (a)$\epsilon_{y}^\text{M2A2}$, (b)$\epsilon_{z}^\text{M2A2}$ and (c)$\hat\epsilon_{z}^\text{M2A2}$ obtained from RTS and derived triangular forms under the unified deflection framework for the Chirp-type POs with non-uniform $n$ BCs.}\label{fig-DeflError_RT-M2A2_chirp}
\end{figure}

For the chirp-type POs, $\epsilon_{(y,z)}^\text{M2A2}(y^*)$ and $\hat\epsilon_{y}^\text{M2A2}(y^*)$ calculated by RTS are plotted in the dash-dot lines in Fig.~\ref{fig-DeflError_RT-M2A2_chirp} (Note $\hat\epsilon_{z}^\text{M2A2}$ does not exist due to $\varepsilon_{z}^\text{M2A2}=0$). The spatial frequencies of $\epsilon_{(y,z)}^\text{M2A2}$ resemble those of $\nabla n$ for the chirp-type POs and minimized for $y^* =0$. Moreover, the relative error $\hat{\epsilon}_{y}^\text{M2A2}$ exhibits an overall parabola trend symmetric with respect to $y^* = 0$. Note that the global parabolic trend observed for $\hat\epsilon_{y}^\text{M2A2}$ with non-uniform $n$ BCs resembles the one observed for $\epsilon_{z}^\text{M1A1}$ with uniform $n$ BCs. However, the underlying mechanism should be distinct according to the analysis in Sec.\ref{sec_M1A1}, since the $n$ distribution along the optical axis is uniform for the non-uniform $n$ BCs. 

To reveal the mechanism behind the parabolic global trend for $\hat\epsilon_{z}^\text{M2A2}$, the triangular form of $\epsilon_{y}^\text{M2A2}$ with non-uniform $n$ BCs is firstly derived under the unified deflection framework as,
\begin{equation}
\epsilon_{y}^\text{M2A2} \approx \varepsilon_y^\text{M1A1} - \varepsilon_y^\text{M2A2} \approx (n_{out} \sin \beta_{out} - \hat{n}_{in} \sin \hat{\beta}_{in}) - n_0 (\sin \beta_{out} - \sin  \hat{\beta}_{in})
\label{eq_DeflError_RT-M2_varep1}.
\end{equation}
The first ‘$\approx$’ sign holds since $\varepsilon_{y}^\text{RTS}$ originally within the definition of $\epsilon_{y}^\text{M2A2}$ is replaced with $\varepsilon_y^\text{M1A1}$, which is physically sound considering the trivial magnitude of $\epsilon_y^\text{M1A1}$ explored in this paper. The second ‘$\approx$’ sign is originating from replacing ${\beta}_{in}$ with $\hat{\beta}_{in}$ assuming the fixed deflection plane.

Introduce the variable $\beta_{out} = \beta_0 + \delta$ and $ \hat{\beta}_{in} = \beta_0 - \delta$, where $\beta_0$ is the angle between the red reference ray and the z-axis as shown in Fig.~\ref{fig-BOS-Schematic}, and $\delta$ is a small angle associated with weak deflections. Further consider that the PO can be treated as a thin deflection interface where Snell's law $ \hat{n}_{in} \cos  \hat{\beta}_{in} = n_{out} \cos \beta_{out}$ applies, Eq. ~(\ref{eq_DeflError_RT-M2_varep1}) can be simplified as,
\begin{equation}
\epsilon_{y}^\text{M2A2}\approx \frac{n_{out} -  \hat{n}_{in}}{\sin \beta_0} \left[ 1 - \frac{2 n_0}{n_{out} + \hat{n}_{in}} \cos^2 \beta_0 \right]\approx (n_{out} - \hat{n}_{in}) \sin\beta_0\approx (n_{out} - \hat{n}_{in}) \beta_0
\label{eq_DeflError_RT-M2_varep2}.
\end{equation}
This expression indicates that $\epsilon_{y}^\text{M2A2}$ is mainly influenced by the non-uniform $n$ BCs via $n_{out} \pm \hat{n}_{in}$, and the angle $\beta_0$. The validity of Eq.~(\ref{eq_DeflError_RT-M2_varep2}) is confirmed by Fig.~\ref{fig-DeflError_RT-M2A2_chirp}(a), within which the gray line plotted according to Eq.~(\ref{eq_DeflError_RT-M2_varep2}) matches the $\epsilon_{y}^\text{M2A2}$ calculated from RTS well. $\epsilon_{y}^\text{M2A2}$'s spatially oscillating feature of varying frequencies can be explained by the term ($n_{out} - \hat{n}_{in}$) with chirp-type fluctuations as confirmed in Fig.~\ref{fig-DeflError_RT-M2A2_chirp}(b). It is noteworthy that the $n_{out} - \hat{n}_{in}$ cause of the chirp-type fluctuations in $\epsilon_{y}^\text{M2A2}$ is distinctive from the one in $\epsilon_{y}^\text{M1A1}$ with non-uniform BCs, which is attributed to the $\Delta\beta_{in}$ according to Eq.~(\ref{eq_DeflError_RT-M1_NUBC_y2}). A similar derivation for $\epsilon_{z}^\text{M2A2}$ can also be made in the $z$-direction as:
\begin{equation}
\epsilon_{z}^\text{M2A2} \approx (n_{out}- \hat{n}_{in})\cos \beta_{0}\approx n_{out}- \hat{n}_{in}.
\label{eq_DeflError_RT-M2_varepz}
\end{equation}

As shown in Fig.~\ref{fig-DeflError_RT-M2A2_chirp}(b), the gray line plotted according to Eq.~(\ref{eq_DeflError_RT-M2_varepz}) also matches the $\epsilon_{z}^\text{M2A2}$ from RTS as expected. Note that since the $\cos \beta_0$ is very close to one, the non-uniform BC term $n_{out} - \hat{n}_{in}$ dominates $\epsilon_{z}^\text{M2A2}$ as confirmed in Fig.~\ref{fig-DeflError_RT-M2A2_chirp}(b). Moreover, since $n_{out} - \hat{n}_{in}$ is minimum for $y^*=0$ for both $\epsilon_{(y,z)}^\text{M2A2}$, thus explains why the minimum of $\epsilon_{(y,z)}^\text{M2A2}$ occurs also there. 

Compared to ${\epsilon}_{y}^\text{M2A2}$, ${\epsilon}_{z}^\text{M2A2}$ yields an much larger amplitude. This amplitude difference is distinct from the M1A1 case as shown in Fig.~\ref{fig-DeflError_RT-M1A1_chirp}(a) and (c), within which ${\epsilon}_{z}^\text{M1A1}$ is much smaller than ${\epsilon}_{y}^\text{M1A1}$. Such observations can be explained straightforwardly within the unified deflection framework, as shown by the above derivations. For M1A1, ${\epsilon}_{y}^\text{M1A1}$ is proportional to $\cos\beta_{in}$ according to Eq.~(\ref{eq_DeflError_RT-M1_NUBC_y2}) while ${\epsilon}_{z}^\text{M1A1}$ relies on $\sin\beta_{in}$ according to Eq.~(\ref{eq_DeflError_RT-M1_NUBC_z}). Thus, for a small $\beta_{in}$, $\epsilon_{z}^\text{M1A1}<\epsilon_{y}^\text{M1A1}$ holds. For M2A2, ${\epsilon}_{y}^\text{M2A2}$ depends on $\sin\beta_0$ while ${\epsilon}_{z}^\text{M2A2}$ depends on $\cos\beta_0$. For small $\beta_0$, $\epsilon_{z}^\text{M2A2}>\epsilon_{y}^\text{M2A2}$ holds instead. Thus, considering the dominate role of $\epsilon_{z}^\text{M2A2}$, it is essential to provide accurate BCs to minimize it according to Eq.~(\ref{eq_DeflError_RT-M2_varepz}).

Next, we proceed to the derivation for the relative error $\hat\epsilon_{y}^\text{M2A2}$ within the unified deflection framework. Following the definition, we can have,
\begin{equation}
\hat{\epsilon}_{y}^\text{M2A2} = \frac{\epsilon_{y}^\text{M2A2}}{\varepsilon_y^\text{RTS}} \approx \frac{\epsilon_{y}^\text{M2A2}}{\varepsilon_y^\text{M1A1}}
\label{eq_DeflError_RT-M2_y1}
\end{equation}
Following the Taylor expansion for $\epsilon_{y}^\text{M2A2}$, the denominator $\varepsilon_{y}^\text{M1A1}$ can be simplified as,
\begin{equation}
\varepsilon_y^\text{M1A1}  = (n_{out} -  \hat{n}_{in}) \sin \beta_0 \cos \delta + (n_{out} +  \hat{n}_{in}) \cos \beta_0 \sin\delta
\label{eq_DeflError_RT-M2_y2}
\end{equation}
Consider all the Snell law by treating the PO as a thin deflection interface, Eq.~(\ref{eq_DeflError_RT-M2_y2}) can be further simplified as,
\begin{equation}
\varepsilon_y^\text{M1A1} = \frac{(n_{out} -  \hat{n}_{in}) \cos \delta}{\sin \beta_0}\approx \frac{n_{out} -  \hat{n}_{in}}{\sin \beta_0}
\label{eq_DeflError_RT-M2_y3}
\end{equation}
Plug Eq.~(\ref{eq_DeflError_RT-M2_y3}) and (~\ref{eq_DeflError_RT-M2_varep2}) into Eq.(~\ref{eq_DeflError_RT-M2_y1}), the triangular formula for $\hat{\epsilon}_{y}^\text{M2A2}$ can be derived as,
\begin{equation}
\hat{\epsilon}_{y}^\text{M2A2} \approx 1-2 n_0 \frac{\cos ^2 \beta_0}{\left(n_{o u t}+\hat{n}_{in}\right)}
\label{eq_DeflError_RT-M2H_y1}
\end{equation}
For the chirp-type PO explored here with non-uniform $n$ BCs, fluctuating are inevitably introduced into $\hat{\epsilon}_{y}^\text{M2A2}$ via the term $n_{out} +  \hat{n}_{in}$. This explains why $\hat{\epsilon}_{y}^\text{M2A2}$ exhibits a slight oscillatory behaviors in Fig.~\ref{fig-DeflError_RT-M2A2_chirp}(c). Since $n \approx1$ for most aerodynamic applications, we can safely assume the following relation,
\begin{equation}
\frac{2 n_0}{n_{out} +  \hat{n}_{in}} \approx 1.
\label{eq_DeflError_RT-M2_assump1}
\end{equation}
Owing to the assumption, all the oscillations in $\hat{\epsilon}_{y}^\text{M2A2}$ should get filtered. Consequently, an elegant form of filtered $\hat{\epsilon}_{y}^\text{M2A2}$ can be finally reached as,
\begin{equation}
\hat{\epsilon}_{y}^\text{M2A2} \approx \sin^2 \beta_0
\label{eq_DeflError_RT-M2_y5}.
\end{equation}
Equation~(\ref{eq_DeflError_RT-M2_y5}) clearly reveals that it is mainly the angle $\beta_0$ that determines the amplitude of $\hat{\epsilon}_{y}^\text{M2A2}$. This is exactly the case as confirmed in Fig.~\ref{fig-DeflError_RT-M2A2_chirp}(c), where the global trend of $\hat{\epsilon}_{y}^\text{M2A2}$ obtained by RTS matches $\sin^2 \beta_0$ well. 
%This conclusion indicates that $\hat{\epsilon}_{y}^\text{M2A2}$ can be rectified by accounting for the deviation angle $\beta_0$ from the principal optical axis.

To explore the ${\epsilon}_{(x,y,z)}^\text{M2A2}$ for more realistic flows, turbulent-type POs are again utilized. ${\epsilon}_{(x,y,z)}^\text{M2A2}$ obtained both by RTS and analytical triangular form derived in Appendix.\ref{sec_A1} are shown in Fig.~\ref{fig-Defl_RT-M2_JHU_epsilon}. Nearly perfect match between the two again verifies the validity of the analytical derivations within the unified framework. Moreover, similar to the chirp-type POs, ${\epsilon}_{z}^\text{M2A2}$ of an order of magnitude larger amplitude than ${\epsilon}_{x,y}^\text{M2A2}$ can be observed. 
\begin{figure}[htb!]
\centering
\includegraphics[width=0.55\textwidth]{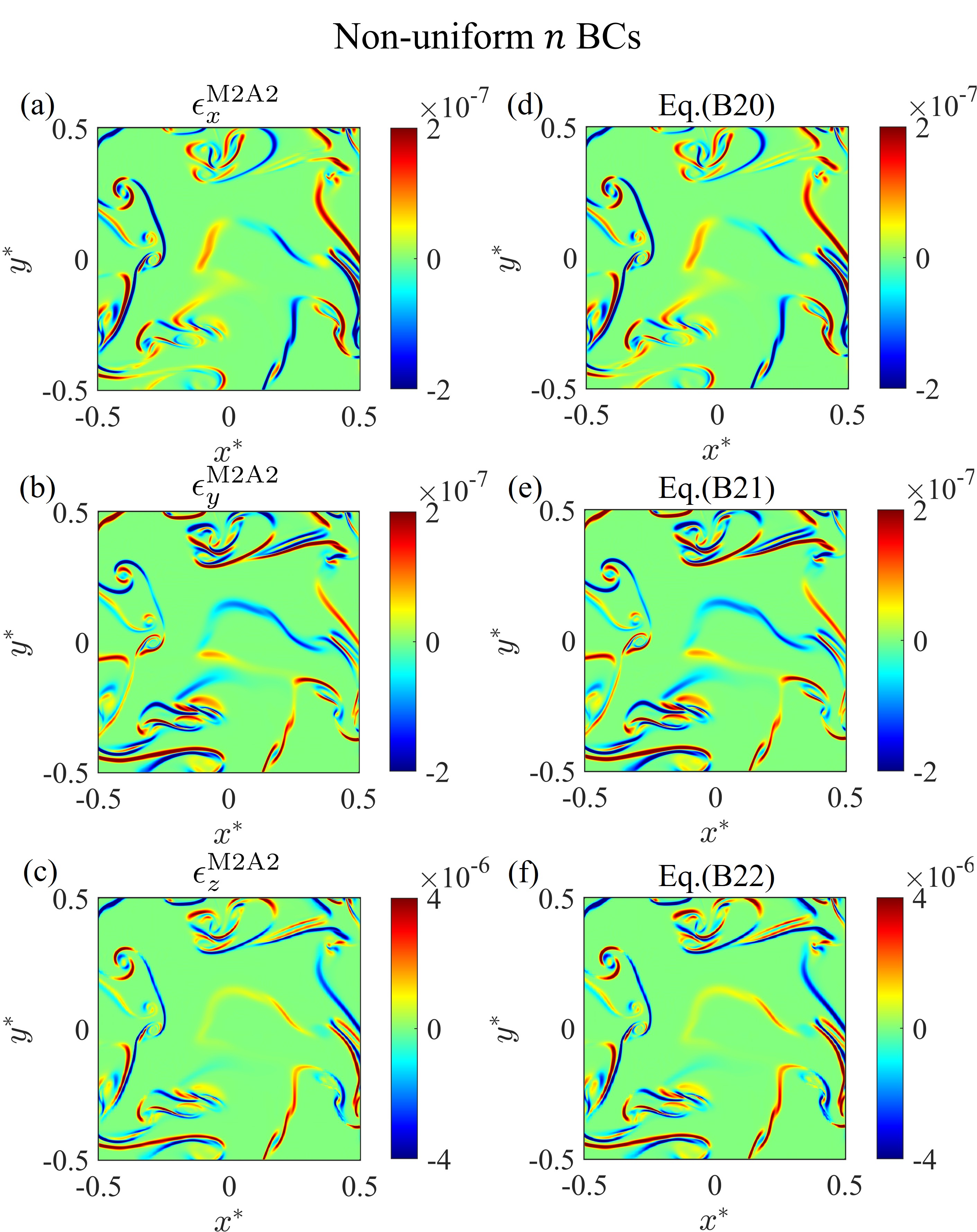}
\caption{Comparisons on $\epsilon_{(x,y,z)}^\text{M2A2}$ obtained from RTS and derived triangular forms under the unified framework(a-c), and theoretical analysis(d-f) for the turbulent-type POs with non-uniform $n$ BCs.}\label{fig-Defl_RT-M2_JHU_epsilon}
\end{figure}

%%%%%%%%%%%%%%%%%%%%%%%%%%%%%%%%%%%%%%%%%%%%%%%%%%%%%
%%%%%%%%%%%%%%%%%%%%%%%%%%%%%%%%%%%%%%%%%%%%%%%%%%%%%

\subsection{M3A4 Method}

\begin{figure}[htb]
\centering
\includegraphics[width=1\textwidth]{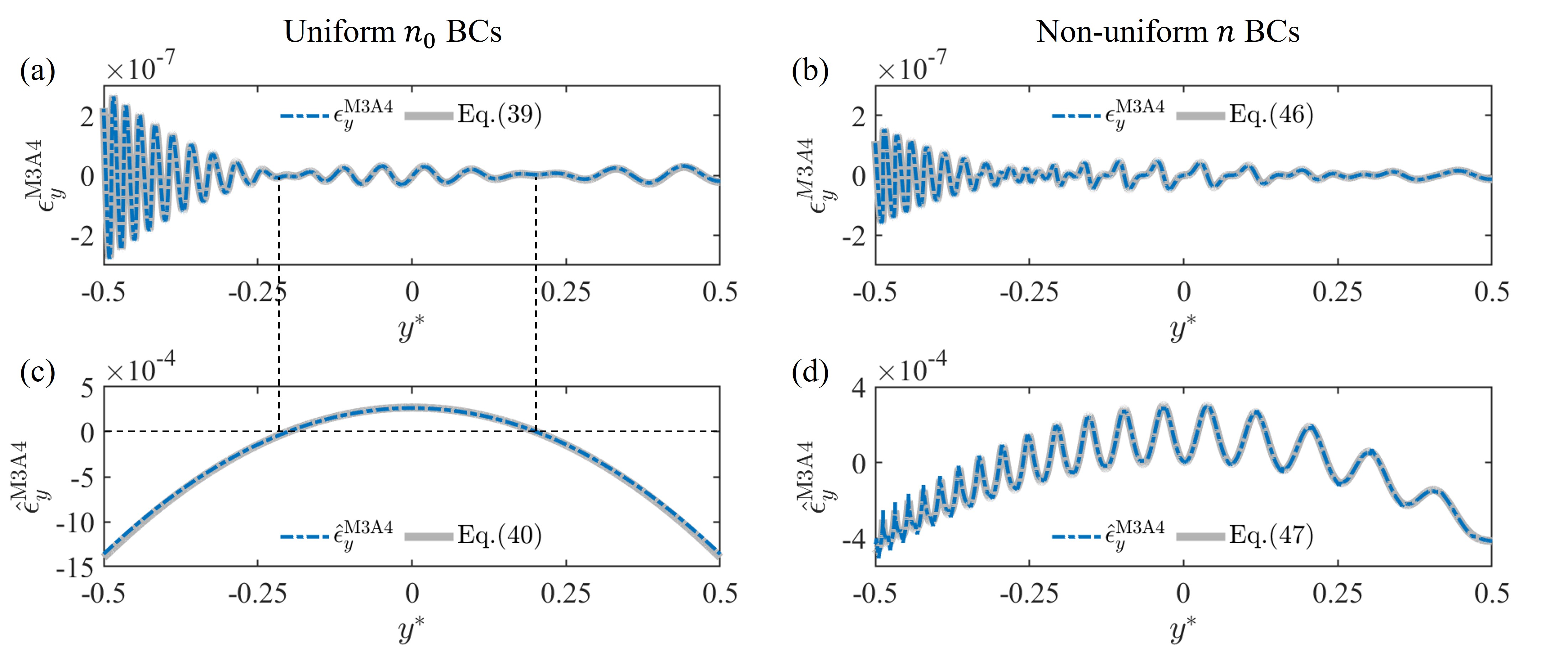}
\caption{Deflection error of the Chirp-type POs obtained with the M3A4 method versus RTS results. (a) Uniform $n_0$ BCs, (b) Non-uniform $n$ BCs.}\label{fig-DeflError_RT-M3A4_chirp}
\end{figure}

Regarding the errors associated with the M3A4 method, $\epsilon_y^\text{M3A4}$ and $\hat{\epsilon}_y^\text{M3A4}$ of the chirp-type POs under uniform $n_0$ BCs are firstly plotted in Fig.~\ref{fig-DeflError_RT-M3A4_chirp}(a) ($\epsilon_z^\text{M3A4}=0$). Compared to the $\epsilon_y^\text{M1A1}$ with uniform $n_0$ BCs in Fig.~\ref{fig-DeflError_RT-M1A1_chirp}(a), global linear/parabolic trends of $\epsilon_y^\text{M3A4}$ vanishes. This is because the global-trend-related errors due to the Gaussian distribution of $n$ along the optical axis are of two orders of magnitude smaller compared to $\epsilon_y^\text{M3A4}$. Moreover, the oscillation amplitude envelope of $\epsilon_y^\text{M3A4}$ evolves in a distinct way from $\epsilon_y^\text{M1A1}$. For example, the oscillation amplitudes of $\epsilon_y^\text{M1A1}$ vanish as $y^* \rightarrow -0.5$ while that of $\epsilon_y^\text{M3A4}$ is not. In addition, $\epsilon_y^\text{M1A1}$ vanishes near $y^* \approx 0$ while the envelope of $\epsilon_y^\text{M3A4}$ has multiple zero points. Regarding the relative error, both the global trend of $\hat{\epsilon}_{y}^\text{M3A4}$ and $\hat{\epsilon}_{y}^\text{M1A1}$ exhibit a parabola symmetry with respect to $|y^*| = 0$, although the parabola opening direction is opposite. To explain these features, the triangular form of $\epsilon_{y}^\text{M3A4}$ with uniform $n_0$ BCs is derived under the unified deflection framework as,
\begin{equation}
\epsilon_y^\text{M3A4} \approx \varepsilon_y^\text{M1A1} - \varepsilon_y^\text{M3A4}  = \varepsilon_y^\text{M1A1} \left[ 1 - \frac{\tan\beta_{out} - \tan \hat{\beta}_{in}}{n_0 (\sin\beta_{out} - \sin \hat{\beta}_{in})} \right]
\label{eq_DeflError_RT-M3_UBC_1}
\end{equation}
Again assuming $\beta_{out} = \beta_0 + \delta$ and $ \hat{\beta}_{in} = \beta_0 - \delta$ and applying Taylor expansion, Eq.~(\ref{eq_DeflError_RT-M3_UBC_1}) simplifies to,
\begin{equation}
\epsilon_y^\text{M3A4} \approx \varepsilon_y^\text{M1A1} \left[ 1 - \frac{2\delta \tan'\beta_0}{2n_0 \delta \sin'\beta_0} \right] \approx \varepsilon_y^\text{M1A1} \left[ 1 - \frac{1}{n_0 \cos^3\beta_0} \right]
\label{eq_DeflError_RT-M3_UBC_2}
\end{equation}
This expression indicates that $\epsilon_y^\text{M3A4}$ is closely linked to $\varepsilon_y^\text{M1A1}$ and only differs by a scaling factor. Consequently, the relative error is given by:
\begin{equation}
\hat{\epsilon}_y^\text{M3A4} \approx \frac{\epsilon_y^\text{M3A4}}{\varepsilon_y^\text{M1A1}} = 1 - \frac{1}{n_0 \cos^3\beta_0}
\label{eq_DeflError_RT-M3_UBC_3}
\end{equation}

The analytical results from Eqs.~(\ref{eq_DeflError_RT-M3_UBC_2}) and (\ref{eq_DeflError_RT-M3_UBC_3}) are in excellent agreement with the M3A4 errors as shown in Fig.~\ref{fig-DeflError_RT-M3A4_chirp}(a) and (c), confirming that $\hat{\epsilon}_y^\text{M3A4}$ with uniform BCs is mainly governed by the off-principal-axis angle $\beta_0$. In addition, a vanished oscillating envelope at two $y^*$ locations can be identified as shown by the markers in Fig.~\ref{fig-DeflError_RT-M3A4_chirp}(a). These locations can be determined by the equation $n_0 \cos^3\beta_0 = 1$, as indicated by the crossing points between the gray dashed lines and the parabola as in Fig.~\ref{fig-DeflError_RT-M3A4_chirp}(c).

The errors associated with the M3A4 method applied to chirp-type POs under non-uniform BCs are shown in Fig.~\ref{fig-DeflError_RT-M3A4_chirp}(b). While the distribution trend is similar to that in Fig.~\ref{fig-DeflError_RT-M3A4_chirp}(a), the numerical ranges differ significantly. Furthermore, the relative error $\hat{\epsilon}_y^{M3A4}$ under non-uniform BCs (Fig.~\ref{fig-DeflError_RT-M3A4_chirp}b) exhibits oscillations, indicating a distinct error mode compared to the case of uniform BCs. We derive the error expression for non-uniform BCs within the unified framework:
\begin{equation}
\epsilon_y^\text{M3A4} \approx \varepsilon_y^\text{M1A1} - \varepsilon_y^\text{M3A4} = \frac{(n_{out} -  \hat{n}_{in})}{\sin\beta_0} - (\tan\beta_{out} - \tan \hat{\beta}_{in})
\label{eq_DeflError_RT-M3_NUBC_1}
\end{equation}
Using the central difference approximation, Eq.~(\ref{eq_DeflError_RT-M3_NUBC_1}) becomes:
\begin{equation}
\epsilon_y^\text{M3A4} \approx \frac{(n_{out} -  \hat{n}_{in})}{\sin\beta_0} - \frac{2\delta}{\cos^2\beta_0}
\label{eq_DeflError_RT-M3_NUBC_2}
\end{equation}
According to Snell’s law, assuming $n_{in} \cos\beta_{in} = n_{out} \cos\beta_{out}$ and applying the small deflection angle $\delta$ hypothesis, the following relationships are obtained:
\begin{equation}
n_{in} (\cos\beta_0 \cos\delta + \sin\beta_0 \sin\delta) = n_{out} (\cos\beta_0 \cos\delta - \sin\beta_0 \sin\delta)
\label{eq_SnellApprox_1}
\end{equation}
\begin{equation}
\delta \approx \tan\delta = \frac{\sin\delta}{\cos\delta} = \frac{(n_{out} -  \hat{n}_{in})\cos\beta_0}{(n_{out} +  \hat{n}_{in})\sin\beta_0}
\label{eq_SnellApprox_2}
\end{equation}

Substituting Eq.~(\ref{eq_SnellApprox_2}) into Eq.~(\ref{eq_DeflError_RT-M3_NUBC_2}) yields the analytical expressions for the absolute and relative errors:
\begin{equation}
\epsilon_y^\text{M3A4} \approx \frac{(n_{out} -  \hat{n}_{in})}{ (n_{out} +  \hat{n}_{in})} \left[ \frac{(n_{out} +  \hat{n}_{in}) \cos\beta_0 - 2}{\sin\beta_0 \cos\beta_0} \right]
\label{eq_DeflError_RT-M3_NUBC_3}
\end{equation}
\begin{equation}
\hat{\epsilon}_y^\text{M3A4} \approx 1 - \frac{2}{(n_{out} +  \hat{n}_{in}) \cos\beta_0}
\label{eq_DeflError_RT-M3_NUBC_4}
\end{equation}

As shown in Fig.~\ref{fig-DeflError_RT-M3A4_chirp}(b), the analytical error expressions Eqs.~(\ref{eq_DeflError_RT-M3_NUBC_3}) and (\ref{eq_DeflError_RT-M3_NUBC_4}) for M3A4 align well with the simulation results, suggesting that under non-uniform BCs, the refractive index difference at the boundaries is a significant error source in addition to the viewing off-axis angle $\beta_0$. Notably, as shown in Eq.~(\ref{eq_DeflError_RT-M3_NUBC_4}), the oscillation of $(n_{out} + n_{in})$ is responsible for the fluctuations in the relative error (Fig.~\ref{fig-DeflError_RT-M3A4_chirp}b).

\begin{figure}[!htb]
\centering
\includegraphics[width=1\textwidth]{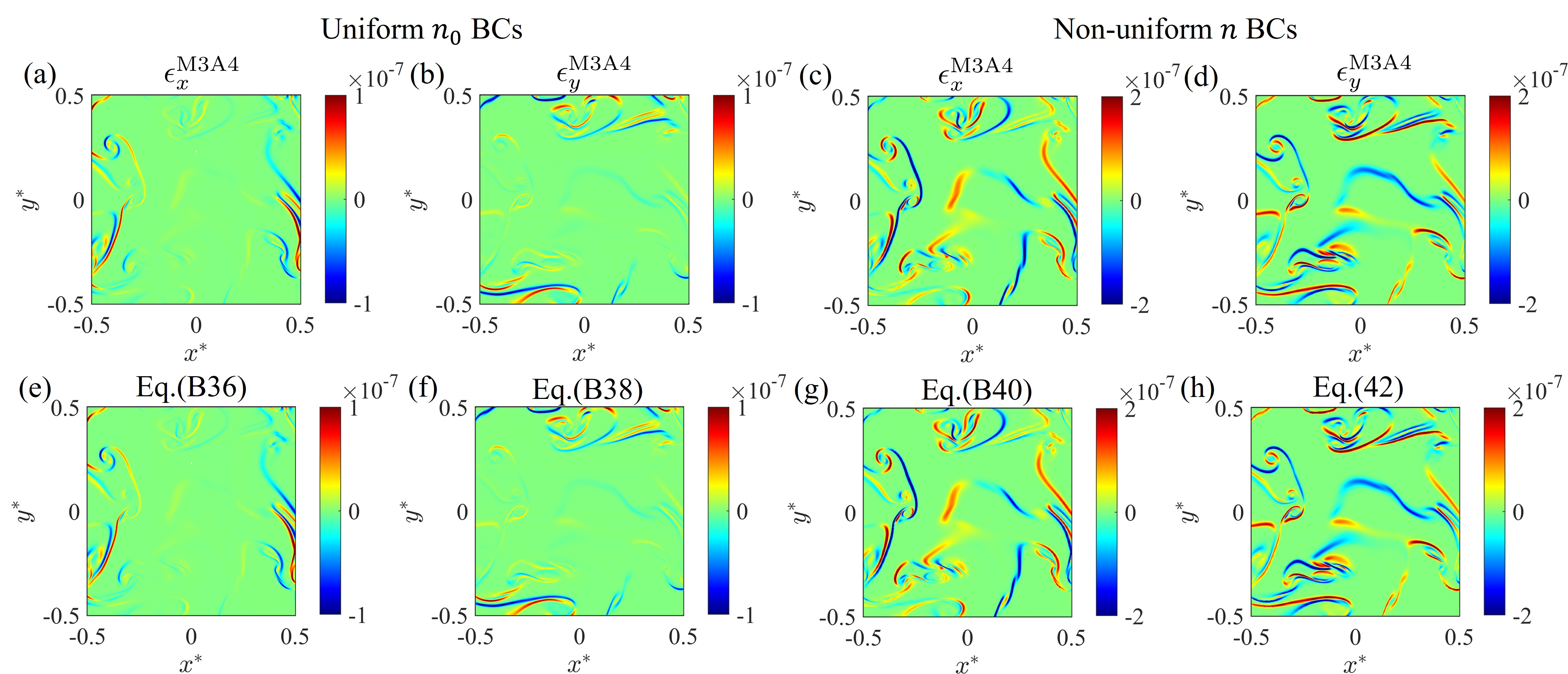}
\caption{Comparisons on $\epsilon_{(x,y)}^\text{M3A4}$ obtained from RTS and derived triangular forms under the unified framework(a-d), and theoretical analysis(e-h) for the turbulent-type POs with uniform $n_0$ BCs and non-uniform $n$ BCs.}\label{fig-Defl_RT-M3_JHU}
\end{figure}

For the turbulence-type POs with uniform BCs, The error distribution of ${\epsilon}_{(x,y)}^\text{M3A4}$ gradually increases as the viewing off-axis angle $\beta_0$ increases (i.e., as $|x^*| \& |y^*| \to 0.5$) as shown in Figs.~\ref{fig-Defl_RT-M3_JHU}(a)--(b), consisted with the observations made for the chirp-type PO in Fig.~\ref{fig-DeflError_RT-M3A4_chirp}(a). Results with non-uniform BCs (Figs.~\ref{fig-Defl_RT-M3_JHU}c--d) also align with the conclusions of Eqs.~(\ref{eq_DeflError_RT-M3_NUBC_3}). Analytical triangular expressions obtained within the unified framework again predict the errors, which match well with the RTS results.

%%%%%%%%%%%%%%%%%%%%%%%%%%%%%%%%%%%%%%%%%%%%%%%%%%%%%

\section{Conclusion}\label{sec_Concl}

Estimating ray deflection from displacement data is the essential step of contemporary two- and three-dimensional BOS. Because the measured PO is unknown a priori and light propagates along nonlinear trajectories, practical schemes inevitably rely on approximations that introduce measurement errors. Grounded in geometric optics, this study unifies the previously disparate ray-deflection estimation methods into a single mathematical structure, the unified deflection estimation framework, and uses it to quantitatively analyze and physically interpret the associated errors.

The central achievement is the framework itself, which integrates methods from the rigorous M1A1 formulation to the highly simplified M3A4 expression into one geometric-optics architecture and clarifies the role of each approximation, namely the thin PO, the uniform boundary refractive index, the paraxial condition, and the perpendicularity between the deflection vector and the optical axis. By expressing every method through the same boundary refractive indices and ray-orientation angles, this approach renders schemes that were historically formulated and interpreted differently directly comparable, thereby bridging the long-standing gap between two- and three-dimensional BOS deflection estimation.

Validated against high-fidelity ray-tracing data for idealized chirp and realistic turbulent POs, the framework yields a quantitative, analytically tractable picture of each method's error. M1A1 attains the highest accuracy and faithfully proxies the true deflection under thin-PO conditions, whereas the widely used M2A2 and M3A4 exhibit systematic biases under non-uniform boundaries, dominated by the boundary refractive-index difference ($n_{out} - n_{in}$) and the viewing off-axis angle ($\beta_0$). Even under uniform boundaries, the additional paraxial and perpendicularity approximations in M3A4 introduce a $\beta_0$-governed error that M2A2 avoids; M2A2 is therefore recommended when boundary refractive indices are unavailable.

Collectively, these results equip the BOS community with explicit, physically grounded error expressions, enabling informed method selection, guiding experimental design such as minimizing off-axis viewing angles and enforcing uniform refractive-index boundaries, and improving tomographic reconstruction as BOS advances toward high-precision, three-dimensional, time-resolved diagnostics. They also provide a natural foundation for uncertainty quantification. Challenges remain for thick, strongly refracting three-dimensional POs with highly nonlinear ray paths; we leave these to future work. Ultimately, this study provides a robust theoretical foundation for high-precision quantitative BOS diagnostics.

\bmhead{Acknowledgements}
This work is supported by the National Natural Science Foundation of China (NSFC Grant No.12572320, No.12502329), the Ningbo Yongjiang Talent Programme (Grant No. 2023A-359-G), and the Fundamental Research Funds for the Central Universities.

\bmhead{Data availability}
Data underlying the results presented in this paper are not publicly available at this time but can be obtained from the authors upon reasonable request.

%\bmhead{Author contribution}
%\textcolor{red}{Conceptualization was done by YX; data generation and acquisition were done by XL; JL and XL performed the investigation, formal analysis, methodology, and visualization and wrote the original draft; all authors contributed to review and editing; YX worked in supervision; and YX and JL helped in funding acquisition.}

\begin{appendices}

\section{Unified Deflection Estimation in 3D Space}\label{sec_A1}

Based on the theoretical foundation discussed in Section~\ref{subsec: Deflection_estimation}, we extend the Unified Deflection Estimation method to broader 3D spatial deflection estimation. To this end, it is necessary to first define the angular notation that differs from the 2D spatial deflection estimation (as shown in Figs.~\ref{fig-BOS-Schematic} and \ref{fig-DeflGeo}): $\alpha$, $\beta$, and $\gamma$ are the angles between the ray vector and the $X$-, $Y$-, and $Z$-axes of the Cartesian coordinate system, respectively. Therefore, deflection in 3D space is defined as:
\begin{equation}
\bm\varepsilon = n_{out}\bm{d}_{out}-n_{in}\bm{d}_{in},
\label{eqA_epsilonDefine}
\end{equation}
and its components are respectively defined as:
\begin{equation}
\begin{aligned}
\varepsilon_x = \bm{i}\cdot\bm\varepsilon = n_{out}\cos{\alpha_{out}}-n_{in}\cos{\alpha}_{in}, \\
\varepsilon_y = \bm{j}\cdot\bm\varepsilon = n_{out}\cos{\beta_{out}}-n_{in}\cos{\beta}_{in}, \\
\varepsilon_z = \bm{k}\cdot\bm\varepsilon = n_{out}\cos{\gamma_{out}}-n_{in}\cos{\gamma}_{in}, \\
\end{aligned}
\label{eqA_epsilonDefine-xyz}
\end{equation}
where, $\bm{i}$, $\bm{j}$, and $\bm{k}$ represent the unit vector along the $x$-, $y$-, and $z$-directions. 

For the M1A1 method, the ray entering the phase object is approximated/estimated, so its estimated deflection vector and its component expressed in trigonometric functions (taking the $x$-direction as an example) are respectively:
\begin{equation}
\bm\varepsilon^{\text{M1A1}} = n_{out}\bm{d}_{out}-\hat{n}_{in}\hat{\bm{d}}_{in},
\label{eqA_M1-define}
\end{equation}
\begin{equation}
\varepsilon^{\text{M1A1}}_x = n_{out}\cos{\alpha_{out}}-\hat{n}_{in}\cos{\hat{\alpha}_{in}}
\label{eqA_M1-define-x}
\end{equation}

The M2A2 method introduces the assumption of constant boundary refractive index based on the M1A1 method, i.e., $n_{out} = n_{in} = n_0$. Therefore, the estimated deflection vector and its trigonometric function representation components (taking the $x$-direction as an example) are respectively
\begin{equation}
\bm\varepsilon^{\text{M2A2}} = n_{0}(\bm{d}_{out}-\hat{\bm{d}}_{in}),
\label{eqA_M2-define}
\end{equation}
\begin{equation}
\varepsilon^{\text{M2A2}}_x = n_{0}(\cos{\alpha_{out}}-\cos{\hat{\alpha}_{in}})
\label{eqA_M2-define-x}
\end{equation}

The definition of the M3A4 method is consistent in both 2D and 3D spaces, and it can estimate deflections only in the $x$- and $y$-directions. Therefore, the vector form it defines is:
\begin{equation}
\bm\varepsilon^{\text{M3A4}} = [\Delta'_x/Z_d,\Delta'_y/Z_d,0],
\label{eqA_M3-define}
\end{equation}
For this vector, its trigonometric components are:
\begin{equation}
\begin{aligned}
\varepsilon^{\text{M3A4}}_x = \frac{\cos{\alpha_{out}}}{\cos{\gamma_{out}}} - \frac{\cos{\hat{\alpha}_{in}}}{\cos{\hat{\gamma}_{in}}}, \\
\varepsilon^{\text{M3A4}}_y = \frac{\cos{\beta_{out}}}{\cos{\gamma_{out}}} - \frac{\cos{\hat{\beta}_{in}}}{\cos{\hat{\gamma}_{in}}},
\end{aligned}
\label{eqA_M3-define-xy}
\end{equation}

In summary, in 3D space, the Theoretical definition of light-ray deflection and the definitions of three deflection estimation methods within a unified framework are shown in Table~\ref{tableA_unified-3D_table}.

%\begin{table}[!htb]
%\caption{Theoretical definition of light ray deflection and three deflection estimation methods under a unified framework ($\varepsilon_y$ as an example for the triangular form).}
%\label{tableA_unified-3D_table}
%\centering
%\begin{tabular}{ccc}
%\hline 
%\text{Methods} &  \text{Vector Form} & \text{Triangular Form} \\
%\hline 
%\text{Theoretical} & $n_{\text{out}} \bm{d}_{\text{out}} - n_{\text{in}} \bm{d}_{\text{in}}$ & $n_{\text{out}} \cos \alpha_{\text{out}} - n_{\text{in}} \cos \alpha_{\text{in}}$ \\
%\text{M1A1} & $n_{\text{out}} \bm{d}_{\text{out}} - \hat{n}_{\text{in}} \hat{\bm{d}}_{\text{in}}$ & $n_{\text{out}} \cos \alpha_{\text{out}} - \hat{n}_{\text{in}} \cos \hat{\alpha}_{\text{in}}$ \\
%\text{M2A2} & $n_0 \bm{d}_{\text{out}} - n_0 \hat{\bm{d}}_{\text{in}}$ & $n_0 \cos \alpha_{\text{out}} - n_0 \cos \hat{\alpha}_{\text{in}}$ \\
%\text{M3A4} & $\left[\Delta_x^{\prime} / Z_d, \Delta_y^{\prime} / Z_d, 0\right]$ & $\frac{\cos \alpha_\text{out}}{\cos \gamma_\text{out}} - \frac{\cos \hat{\alpha}_\text{in}}{\cos \hat{\gamma}_\text{in}}$ \\
%\hline
%\end{tabular}
%\end{table}

\begin{table}[!htb]
\centering
\caption{Theoretical definition of light ray deflection and three deflection estimation methods under a unified framework.}
\label{tableA_unified-3D_table}
\begin{tabular}{cccc}
\toprule
\multirow{2}{*}{\text{Methods}} & \multicolumn{3}{c}{\text{Triangular Form}} \\
\cmidrule(lr){2-4}
& $x$-direction & $y$-direction & $z$-direction \\
\midrule
Theoretical 
    & $n_{\text{out}} \cos \alpha_{\text{out}} - n_{\text{in}} \cos \alpha_{\text{in}}$ 
    & $n_{\text{out}} \cos \beta_{\text{out}} - n_{\text{in}} \cos \beta_{\text{in}}$ 
    & $n_{\text{out}} \cos \gamma_{\text{out}} - n_{\text{in}} \cos \gamma_{\text{in}}$ \\
M1A1 
    & $n_{\text{out}} \cos \alpha_{\text{out}} - \hat{n}_{\text{in}} \cos \hat{\alpha}_{\text{in}}$ 
    & $n_{\text{out}} \cos \beta_{\text{out}} - \hat{n}_{\text{in}} \cos \hat{\beta}_{\text{in}}$ 
    & $n_{\text{out}} \cos \gamma_{\text{out}} - \hat{n}_{\text{in}} \cos \hat{\beta}_{\text{in}}$  \\
M2A2 
    & $n_0 ( \cos \alpha_{\text{out}} - \cos \hat{\alpha}_{\text{in}} )$ 
    & $n_0 ( \cos \beta_{\text{out}} - \cos \hat{\beta}_{\text{in}} )$ 
    & $n_0 ( \cos \gamma_{\text{out}} - \cos \hat{\gamma}_{\text{in}} )$  \\
M3A4 
    & $\frac{\cos \alpha_\text{out}}{\cos \gamma_\text{out}} - \frac{\cos \hat{\alpha}_\text{in}}{\cos \hat{\gamma}_\text{in}}$ 
    & $\frac{\cos \beta_\text{out}}{\cos \gamma_\text{out}} - \frac{\cos \hat{\beta}_\text{in}}{\cos \hat{\gamma}_\text{in}}$ 
    & 0  \\
\bottomrule
\end{tabular}
\end{table}

\section{Analytical Errors of Deflection Estimation in 3D Space}\label{sec_A2}

\subsection{M1A1 Method}

\subsubsection{Uniform $n_0$ BCs}

The M1A1 method uses the fewest assumptions, so its error for the theoretical deflection is relatively small. Under the thin phase object assumption, it can be considered that the deviation between the estimated $\hat{\bm{d}}_{in}$ and the true $\bm{d}_{in}$ is small, meaning the angles between them ($\Delta\alpha = \hat{\alpha}_{in}-\alpha_{in}$, $\Delta\beta = \hat{\beta}_{in}-\beta_{in}$, and $\Delta\gamma = \hat{\gamma}_{in}-\gamma_{in}$) is small. Therefore, under uniform $n_0$ BCs ($n_{in} = n_{out} = n_{0}$), the analytical expression for the error of the M1A1 method within the unified representation framework is:
\begin{equation}
\begin{aligned}
\epsilon^{\text{M1A1}}_x
  &= n_0(\cos\alpha_{out}-\cos\alpha_{in})
   - n_0(\cos\alpha_{out}-\cos\hat\alpha_{in})  \\
  &= n_0(\cos\hat\alpha_{in}-\cos\alpha_{in}) \\
  &= n_0\bigl[\cos(\alpha_{in}+\Delta\alpha)-\cos\alpha_{in}\bigr]
\end{aligned}
\label{eqA_error-M1-uBC-3D-x}
\end{equation}
Using the principle of prior approximate differentiation, the above formula is approximated as
\begin{equation}
\epsilon^{\text{M1A1}}_x \approx n_0 \Delta \alpha \cos'\alpha_{in} = -\,n_0 \Delta\alpha \sin\beta_{in}
\label{eqA_error-M1-uBC-3D-x2}
\end{equation}
Similarly,
\begin{equation}
\epsilon^{\text{M1A1}}_y = -\,n_0 \Delta\beta \sin\beta_{in}
\label{eqA_error-M1-uBC-3D-y}
\end{equation}
\begin{equation}
\epsilon^{\text{M1A1}}_z = -\,n_0 \Delta\gamma \sin\gamma_{in}
\label{eqA_error-M1-uBC-3D-z}
\end{equation}

\subsubsection{Non-uniform $n$ BCs}

For non-uniform $n$ BCs ($n_{in} \neq n_{out}$), the analytical expression of the M1A1 method error under the unified representation framework is:
\begin{equation}
\begin{aligned}
\epsilon_x
  &= (n_{out}\cos\alpha_{out}-n_{in}\cos\alpha_{in}) - (n_{out}\cos\alpha_{out}-\hat{n}_{in}\cos\hat\alpha_{in}) \\
  &= \hat{n}_{in}\cos\hat\alpha_{in} - n_{in}\cos\alpha_{in} \\
  &= n_{in}(\frac{\hat{n}_{in}}{n_{in}}\cos{(\alpha_{in}+\Delta \alpha)}-\cos\alpha_{in})
\end{aligned}
\label{eqA_error-M1-nBC-3D-x1}
\end{equation}
Under the thin phase object approximation, it can be assumed that $\hat{n}_{in} / n_{in} \approx 1$. Then, using the principle of prior approximate differentiation, the above expression is approximated as
\begin{equation}
\epsilon^{\text{M1A1}}_x \approx n_{in} \Delta \alpha \cos'\alpha_{in} = -\,n_{in} \Delta\alpha \sin\alpha_{in}
\label{eqA_error-M1-nBC-3D-x2}
\end{equation}
Similarly, it can be obtained that
\begin{equation}
\epsilon^{\text{M1A1}}_y = -\,n_{in} \Delta\beta \sin\beta_{in}
\label{eqA_error-M1-nBC-3D-y}
\end{equation}
\begin{equation}
\epsilon^{\text{M1A1}}_z = -\,n_{in} \Delta\gamma \sin\gamma_{in}
\label{eqA_error-M1-nBC-3D-z}
\end{equation}

\subsection{M2A2 Method}

Under uniform $n_0$ BCs, since $n_{in} = n_{out} = n_0$, the M2A2 method retreats to the M1A1 method, and the error at this point is equivalent to Eqs.~(\ref{eqA_error-M1-uBC-3D-x2})--(\ref{eqA_error-M1-uBC-3D-z}).

For non-uniform $n$ BCs, assuming the ray deflection angle satisfies the following condition:
\begin{equation}
\begin{aligned}
&\alpha_{out} = \alpha_0+\delta_{\alpha}; &\hat{\alpha}_{in} = \alpha_0-\delta_{\alpha}; \\
&\beta_{out} = \beta_0+\delta_{\beta}; &\hat{\beta}_{in} = \beta_0-\delta_{\beta}; \\
&\gamma_{out} = \gamma_0+\delta_{\gamma}; &\hat{\gamma}_{in} = \gamma_0-\delta_{\gamma}; \\
\end{aligned}
\label{eqA_error-M2-nBC-3D-SmallAngle}
\end{equation}
where $\alpha_0$, $\beta_0$, and $\gamma_0$ represent the angles between the undeflected reference ray and the coordinate axes; $\delta_\alpha$, $\delta_\beta$, $\delta_\gamma$ are small deflection angles. At this point, $\varepsilon^{\text{M1A1}}_x$ and $\varepsilon^{\text{M2A2}}_x$ can be simplified to:
\begin{equation}
\begin{aligned}
\varepsilon^{\text{M1A1}}_x
& = n_{out}\cos{\alpha_{out}} - \hat{n}_{in}\cos{\hat{\alpha}_{in}} \\
& = n_{out}\cos{(\alpha_{0}+\delta_\alpha)} - \hat{n}_{in}\cos{(\alpha_{0}-\delta_\alpha)} \\
& = (n_{out} - \hat{n}_{in})\cos{\alpha_0}\cos{\delta_\alpha} - (n_{out} + \hat{n}_{in})\sin{\alpha_0}\sin{\delta_\alpha} \\
& \approx (n_{out} - \hat{n}_{in})\cos{\alpha_0} - (n_{out} + \hat{n}_{in})\delta_\alpha \sin{\alpha_0}, 
\end{aligned}
\label{eqA_error-M2-nBC-3D-M1smip}
\end{equation}
\begin{equation}
\begin{aligned}
\varepsilon^{\text{M2A2}}_x
& = n_{0}(\cos{\alpha_{out}} - \cos{\hat{\alpha}_{in}}) \\
& = n_{0}(\cos{(\alpha_{0}+\delta_\alpha)} - \cos{(\alpha_{0}-\delta_\alpha)}) \\
& = -2n_{0}\sin{\alpha_0}\sin{\delta_\alpha}  \\
& \approx -2n_{0} \delta_\alpha \sin{\alpha_0} , 
\end{aligned}
\label{eqA_error-M2-nBC-3D-M2smip}
\end{equation}
The simplification process of Eqs.~(\ref{eqA_error-M2-nBC-3D-M1smip}) and (\ref{eqA_error-M2-nBC-3D-M2smip}) uses the small-angle approximation. Substituting the above two equations into the unified framework expression for the M2A2's error:
\begin{equation}
\begin{aligned}
\epsilon^{\text{M2A2}}_x
& = \varepsilon^{\text{RTS}}_x - \varepsilon^{\text{M2A2}}_x \approx \varepsilon^{\text{M1A1}}_x - \varepsilon^{\text{M2A2}}_x \\
& = (n_{out} - \hat{n}_{in})\cos{\alpha_0} - (n_{out} + \hat{n}_{in})\delta_\alpha \sin{\alpha_0} + 2n_{0} \delta_\alpha \sin{\alpha_0} \\
& \approx (n_{out} - \hat{n}_{in})\cos{\alpha_0}, 
\end{aligned}
\label{eqA_error-M2-nBC-3D-x}
\end{equation}
where $\varepsilon_x^{\text{M1A1}}$ is used as an approximation for the true/ray-traced value $\varepsilon_x^{\text{RTS}}$, and the approximation condition, $2n_0 / (n_{in}+n_{out})=1$, is introduced. The errors in the $y$- and $z$-directions can be derived using the same method:
\begin{equation}
\epsilon^{\text{M2A2}}_y \approx (n_{out} - \hat{n}_{in})\cos{\beta_0}, 
\label{eqA_error-M2-nBC-3D-y}
\end{equation}
\begin{equation}
\epsilon^{\text{M2A2}}_z \approx (n_{out} - \hat{n}_{in})\cos{\gamma_0}, 
\label{eqA_error-M2-nBC-3D-z}
\end{equation}

Based on Eq.~(\ref{eqA_error-M2-nBC-3D-x}), further establish the analytical expression for the relative error of M2A2:
\begin{equation}
\begin{aligned}
\hat{\epsilon}^{\text{M2A2}}_x
& = \frac{\epsilon^{\text{M2A2}}_x}{\varepsilon^{\text{RTS}}_{x}} \approx \frac{\epsilon^{\text{M2A2}}_x}{\varepsilon^{\text{M1A1}}_{x}} \\
& = \frac{(n_{out} - \hat{n}_{in})\cos{\alpha_0}}{(n_{out} - \hat{n}_{in})\cos{\alpha_0} - (n_{out} + \hat{n}_{in})\delta_\alpha \sin{\alpha_0}}
\end{aligned}
\label{eqA_errorR-M2-nBC-3D-x1}
\end{equation}

% To further simplify the above equation, the current light deflection phenomenon (under the approximation of small deflection angles and thin phase object) can be approximated as interface refraction, and the expression for $\delta_\alpha$ is established based on Snell's law:
To further simplify the above equation, the current light deflection phenomenon (under the approximation of small deflection angles and thin phase objects) can be approximated as interface refraction, and the expression for $[\delta_\alpha, \delta_\beta, \delta_\gamma]$ can be established based on Snell's law. Taking the light deflection in the X-direction as an example, the corresponding refractive index gradient is along the x-axis; therefore, the X-direction is approximately the normal direction of the refractive interface, while the Y- and Z-directions are the tangential directions. The two directions are treated separately. For the normal direction, Snell's law in its sine form is satisfied:
\begin{equation}
\begin{aligned}
& n_{in} \sin \alpha_{in} = n_{out} \sin \alpha_{out} \\
& \Rightarrow \quad n_{in} \left( \sin \alpha_0 \cos \delta_\alpha - \cos \alpha_0 \sin \delta_\alpha \right) = n_{out} \left( \sin \alpha_0 \cos \delta_\alpha + \cos \alpha_0 \sin \delta_\alpha \right) \\
& \Rightarrow \quad (n_{in} - n_{out}) \sin \alpha_0 \cos \delta_\alpha = (n_{out} + n_{in}) \cos \alpha_0 \sin \delta_\alpha \\
& \Rightarrow \quad \sin \delta_\alpha = - \frac{(n_{out} - n_{in}) \sin \alpha_0}{(n_{out} + n_{in}) \cos \alpha_0} \cos \delta_\alpha \\
& \Rightarrow \quad \delta_\alpha = - \frac{n_{out} - n_{in}}{n_{out} + n_{in}} \tan \alpha_0
\end{aligned}
\label{eqA_errorR-M2-SnellLaw-alpha}
\end{equation}
For the tangential direction, Snell's law in cosine form is satisfied:
\begin{equation}
\begin{aligned}
& n_{in} \cos \beta_{in} = n_{out} \cos \beta_{out} \\
& \Rightarrow \quad \delta_\beta = \frac{n_{out} - n_{in}}{n_{out} + n_{in}} \cot \beta_0
\end{aligned}
\label{eqA_errorR-M2-SnellLaw-beta}
\end{equation}
\begin{equation}
\begin{aligned}
& n_{in} \cos \gamma_{in} = n_{out} \cos \gamma_{out} \\
& \Rightarrow \quad \delta_\gamma = \frac{n_{out} - n_{in}}{n_{out} + n_{in}} \cot \gamma_0
\end{aligned}
\label{eqA_errorR-M2-SnellLaw-gamma}
\end{equation}

Substituting the result from Eq.~(\ref{eqA_errorR-M2-SnellLaw-alpha}) into Eq.~(\ref{eqA_errorR-M2-nBC-3D-x1}):
\begin{equation}
\hat{\epsilon}^{\text{M2A2}}_x = \cos^{2}{\alpha_0}
\label{eqA_errorR-M2-nBC-3D-x2}
\end{equation}

Using the same derivation, the relative error of the M2A2 method in the $y$-directions can be obtained:
\begin{equation}
\hat{\epsilon}^{\text{M2A2}}_y = \cos^{2}{\beta_0}
\label{eqA_errorR-M2-nBC-3D-y}
\end{equation}

It should be noted that under non-uniform boundary conditions, since $\varepsilon_z^{\text{M1A1}} \approx \varepsilon_z^{\text{RTS}} = 0$, there is no $\hat\epsilon_z^{\text{M2A2}}$.

\subsection{M3A4 Method}

\subsubsection{Uniform $n_0$ BCs}

For uniform $n_0$ BCs, the expression for the error of the M3A4 method within the unified framework is:
\begin{equation}
\epsilon^{\text{M3A4}}_x = \varepsilon^{\text{RTS}}_x - \varepsilon^{\text{M3A4}}_x \approx \varepsilon^{\text{M1A1}}_x - \varepsilon^{\text{M3A4}}_x,
\label{eqA_error-M3-uBC-3D-x1}
\end{equation}
where $\varepsilon^{\text{M1A1}}_x$ is used as the approximation of $\varepsilon^{\text{RTS}}_x$. Next, simplify $\varepsilon^{\text{M3A4}}_x$ accordingly:
\begin{equation}
\varepsilon^{\text{M3A4}}_x = \frac{\cos{\alpha_{out}}}{\cos{\gamma_{out}}} - \frac{\cos{\hat\alpha_{in}}}{\cos{\hat\gamma_{in}}} = \text{term1} - \text{term2},
\label{eqA_error-M3-uBC-3D-epsilon0}
\end{equation}
where,  
\begin{equation}
\begin{aligned}
\text{term1} & = \frac{\cos{\alpha_{out}}}{\cos{\gamma_{out}}} = \frac{\cos(\alpha_0 + \delta_\alpha)}{\cos(\gamma_0 + \delta_\gamma)} \\
& = \frac{\cos(\alpha_0)\cos(\delta_\alpha) - \sin(\alpha_0)\sin(\delta_\alpha)}{\cos(\gamma_0)\cos(\delta_\gamma) - \sin(\gamma_0)\sin(\delta_\gamma)} \\
& \approx \frac{\cos(\alpha_0) - \delta_\alpha \sin(\alpha_0)}{\cos(\gamma_0) - \delta_\gamma \sin(\gamma_0)} \\
& = \frac{\cos(\alpha_0) - \delta_\alpha \sin(\alpha_0)}{\cos(\gamma_0)} \frac{1}{\left(1 - \delta_\gamma \tan(\gamma_0)\right)} \\
& \approx \frac{\cos(\alpha_0) - \delta_\alpha \sin(\alpha_0)}{\cos(\gamma_0)} \left(1 + \delta_\gamma \tan(\gamma_0)\right) \\
& = \frac{\cos(\alpha_0) }{\cos(\gamma_0)} - \frac{\delta_\alpha \sin(\alpha_0) \cos(\gamma_0) - \delta_\gamma \cos(\alpha_0) \sin(\gamma_0)}{\cos^2 \gamma_0}
\end{aligned}
\label{eqA_error-M3-uBC-3D-epsilon1}
\end{equation}
where the small-angle approximation and the relation, $1/(1 - \delta_\gamma \tan\gamma_0) \approx (1 + \delta_\gamma \tan\gamma_0)$, are used. Similarly, it can be derived:
\begin{equation}
\text{term2} = \frac{\cos(\alpha_0) }{\cos(\gamma_0)} + \frac{\delta_\alpha \sin(\alpha_0) \cos(\gamma_0) - \delta_\gamma \cos(\alpha_0) \sin(\gamma_0)}{\cos^2 \gamma_0}
\label{eqA_error-M3-uBC-3D-epsilon2}
\end{equation}

To eliminate the $\delta_\alpha$ and $\delta_\gamma$ present in Eq.~(\ref{eqA_error-M3-uBC-3D-epsilon1}) and (\ref{eqA_error-M3-uBC-3D-epsilon2}), we use a first-order Taylor expansion, i.e., $f(x+\delta)-f(x-\delta) \approx 2\delta f'(x)$, to simplify $\varepsilon^{\text{M1A1}}_x$:
\begin{equation}
\begin{aligned}
\varepsilon^{\text{M1A1}}_x & = n_0(\cos{\alpha_{out}}-\cos{\alpha_{in}}) \\
& = n_0(\cos{(\alpha_{0}+\delta_\alpha)}-\cos{(\alpha_{0}-\delta_\alpha)} ) \\
& = -2n_0 \delta_\alpha \sin\alpha_0
\end{aligned}
\label{eqA_error-M3-uBC-3D-deltaAlpha1}
\end{equation}
and on this basis, establish the relationship between $\delta_\alpha$ and $\varepsilon^{\text{M1A1}}_x$:
\begin{equation}
\delta_\alpha = -\frac{\varepsilon^{\text{M1A1}}_x}{2n_0 \sin\alpha_0},
\label{eqA_error-M3-uBC-3D-deltaAlpha2}
\end{equation}

Similarly, we can obtain:
\begin{equation}
\delta_\beta = -\frac{\varepsilon^{\text{M1A1}}_y}{2n_0 \sin\beta_0},
\label{eqA_error-M3-uBC-3D-deltaBeta}
\end{equation}
\begin{equation}
\delta_\gamma = -\frac{\varepsilon^{\text{M1A1}}_z}{2n_0 \sin\gamma_0},
\label{eqA_error-M3-uBC-3D-deltaGamma}
\end{equation}

Substituting Eqs.~(\ref{eqA_error-M3-uBC-3D-epsilon0})--(\ref{eqA_error-M3-uBC-3D-epsilon2}), (\ref{eqA_error-M3-uBC-3D-deltaAlpha2}), and (\ref{eqA_error-M3-uBC-3D-deltaGamma}) into Eq.~(\ref{eqA_error-M3-uBC-3D-x1}) yields:
\begin{equation}
\epsilon^{\text{M3A4}}_x = \varepsilon^{\text{M1A1}}_x \left [ 1-\frac{1}{n_0 \cos\gamma_0} \right] + \varepsilon^{\text{M1A1}}_z \frac{\cos \alpha_0}{n_0 \cos^2 \gamma_0},
\label{eqA_error-M3-uBC-3D-x2}
\end{equation}
and
\begin{equation}
\hat\epsilon^{\text{M3A4}}_x = \left [ 1-\frac{1}{n_0 \cos\gamma_0} \right] + \frac{\varepsilon^{\text{M1A1}}_z}{\varepsilon^{\text{M1A1}}_x} \frac{\cos \alpha_0}{n_0 \cos^2 \gamma_0}
\label{eqA_errorR-M3-uBC-3D-x}
\end{equation}

Similarly,
\begin{equation}
\epsilon^{\text{M3A4}}_y = \varepsilon^{\text{M1A1}}_y \left [ 1-\frac{1}{n_0 \cos\gamma_0} \right] + \varepsilon^{\text{M1A1}}_z \frac{\cos \beta_0}{n_0 \cos^2 \gamma_0},
\label{eqA_error-M3-uBC-3D-y}
\end{equation}
and
\begin{equation}
\hat\epsilon^{\text{M3A4}}_y = \left [ 1-\frac{1}{n_0 \cos\gamma_0} \right] + \frac{\varepsilon^{\text{M1A1}}_z}{\varepsilon^{\text{M1A1}}_y} \frac{\cos \beta_0}{n_0 \cos^2 \gamma_0}
\label{eqA_errorR-M3-uBC-3D-y}
\end{equation}

\subsubsection{Non-uniform $n$ BCs}
 
Under non-uniform $n$ BCs, the M3A4 error can be analytically expressed by combining Eqs.~(\ref{eqA_error-M2-nBC-3D-M1smip}), (\ref{eqA_errorR-M2-SnellLaw-alpha}), (\ref{eqA_errorR-M2-SnellLaw-gamma}), and (\ref{eqA_error-M3-uBC-3D-epsilon0})--(\ref{eqA_error-M3-uBC-3D-epsilon2}) as: 
\begin{equation}
\begin{aligned}
\epsilon^{\text{M3A4}}_x 
& = \varepsilon^{\text{RTS}}_x - \varepsilon^{\text{M3A4}}_x \approx \varepsilon^{\text{M1A1}}_x - \varepsilon^{\text{M3A4}}_x \\
& = \varepsilon^{\text{M1A1}}_x \left [ 1 - \frac{2}{(n_{out}+n_{in}) \cos \gamma_0} \right ] \\
& = \frac{n_{out} - n_{in}}{\cos \alpha_0} \left [ 1 - \frac{2}{(n_{out}+n_{in}) \cos \gamma_0} \right ],
\end{aligned}
\label{eqA_error-M3-nBC-3D-x}
\end{equation}
and
\begin{equation}
\begin{aligned}
\hat\epsilon^{\text{M3A4}}_x 
& = \frac{\varepsilon^{\text{RTS}}_x - \varepsilon^{\text{M3A4}}_x}{\varepsilon^{\text{RTS}}_x} \approx \frac{\varepsilon^{\text{M1A1}}_x - \varepsilon^{\text{M3A4}}_x}{\varepsilon^{\text{M1A1}}_x}  \\
& = 1 - \frac{2}{(n_{out}+n_{in}) \cos \gamma_0}
\end{aligned}
\label{eqA_errorR-M3-nBC-3D-x}
\end{equation}

If the assumption $2n_0/(n_{in}+n_{out}) \approx 1$ is further introduced, Eq.~(\ref{eqA_errorR-M3-nBC-3D-x}) can be simplified into a more concise form:
\begin{equation}
\begin{aligned}
\hat\epsilon^{\text{M3A4}}_x \approx 1 - \frac{1}{n_{0} \cos \gamma_0}
\end{aligned}
\label{eqA_errorR-M3-nBC-3D-x2}
\end{equation}

Similarly, based on the rotational symmetry of the errors in the $x$- and $y$-directions, the analytical expressions for the error of M3A4 in the $y$-direction are:
\begin{equation}
\begin{aligned}
\epsilon^{\text{M3A4}}_y  = \varepsilon^{\text{M1A1}}_y \left [ 1 - \frac{2}{(n_{out}+n_{in}) \cos \gamma_0} \right ],
\end{aligned}
\label{eqA_error-M3-nBC-3D-y}
\end{equation}
and
\begin{equation}
\begin{aligned}
\hat\epsilon^{\text{M3A4}}_y = 1 - \frac{2}{(n_{out}+n_{in}) \cos \gamma_0}
\end{aligned}
\label{eqA_errorR-M3-nBC-3D-y}
\end{equation}

If the assumption $2n_0/(n_{in}+n_{out}) \approx 1$ is further introduced, Eq.~(\ref{eqA_errorR-M3-nBC-3D-y}) can be simplified into a more concise form:
\begin{equation}
\begin{aligned}
\hat\epsilon^{\text{M3A4}}_y \approx 1 - \frac{1}{n_{0} \cos \gamma_0}
\end{aligned}
\label{eqA_errorR-M3-nBC-3D-y2}
\end{equation}




\end{appendices}


\bibliography{references-zotero-new}% common bib file

\end{document}